%% file: sand.tex
\documentclass{amsart}  
\usepackage{a4wide,amsmath,amsfonts, amssymb, comment, epsfig}     
\usepackage{enumerate}
\usepackage[all]{xy}
\input xy
\xyoption{all}
\theoremstyle{plain}

  \newtheorem{THEO}{Theorem}[section]

  \newtheorem{PROP}{Proposition}[section]

  \newtheorem{lemma}{Lemma}[section]

  \newtheorem{corollary}{Corollary}[section]

\theoremstyle{definition}

  \newtheorem{DEF}{Definition}[section]

  \newtheorem{examp}{Example}[section]

\newenvironment{example}{
  \begin{examp}
                        }
  {\end{examp}    }
  
  \newtheorem{conj}{Conjecture}   

\theoremstyle{remark}

  \newtheorem{remark}{Remark}
    
  \newtheorem{rem}{Remark}[section]
  
\newtheorem{OP}{Open Problem}
   
\newcommand{\para}[1]{\left\langle#1\right\rangle}
\newcommand{\parg}[1]{\left\{#1\right\}}

\newcommand{\cD}{{\mathcal D}} 

\newcommand{\cN}{{\mathcal N}}

\newcommand{\oppE}{\exists}
\newcommand{\oppA}{\forall}
\newcommand{\IR}{{\mathbb R}} 

\newcommand{\incl}{\subseteq}
\def\LP{\par\noindent}
   
\def\MP{\medskip\LP}

\newcommand{\virg}[1]{``#1''}
\def\til{~}

\newcommand{\IN}{{\mathbb N}}
\newcommand{\IZ}{{\mathbb Z}}

\begin{document}
\title[Parallel or spurious sandpile models]{Parallel sandpiles or spurious bidirectional icepiles?\\ \footnotesize{An interesting dilemma from a Formenti--Perrot paper}}
\author{Gianpiero Cattaneo}
\author{Luca Manzoni}
\address{Dipartimento di Informatica, Sistemistica e Comunicazione,\\
  Universit\`a di Milano--Bicocca,\\
  Viale Sarca 336--U14, I--20126 Milano (Italy)
  \and
  Dipartimento di Matematica e Geoscienze,\\
  Università degli Studi di Trieste,\\
  Via Alfonso Valerio 12/1, 34127 Trieste (Italy)
}
\email{cattang@live.it}
\email{lmanzoni@units.it}
\keywords{sandpile model, one-dimensional, parallel update, spurious icepile model}
\begin{abstract}
In a recent paper E. Formenti and K. Perrot (FP) introduce a global rule assumed to describe the discrete time dynamics associated with a sandpile model under the parallel application of a suitable local rule acting on $d$ dimensional lattices of cells equipped with uniform neighborhood.

In this paper we submit this approach to a critical analysis, in the simplest elementary  particular case of a one-dimensional lattice, which can be divided in two parts.

In the first part we prove that the FP global rule does not describe the dynamics of standard sandpiles, but rather furnishes a description of the quite different situation of height difference between consecutive piles. This is a semantic uncorrect difference of interpretation.

In the second part we investigate the consequences of the uncorrect FP assumption proving that their global rule describes a bidirectional spurious dynamics of icepiles (rather than sandpiles), in the sense that this latter is the consequence of application of three local rules: bidirectional vertical rule, bidirectional horizontal rule (typical of icepiles), and a granule jump from the bottom to the top (spurious rule of the dynamics).
\end{abstract}
\maketitle
\input spF-intro
\input spF-part1
\input spF-part2
\input spF-part3
\input spF-part4
\bibliographystyle{amsalpha}   
\bibliography{sand}
\end{document}

%% file: spF-intro.tex
\section{Introduction}
In a recent paper of E. Formenti and K. Perrot (FP) \cite{FP20}, whose title explicitly involves \virg{sandpiles on lattices}, it is formalised a
$d$--dimensional lattice of \emph{cells} $\IZ^d$ on which \emph{configurations} are defined as mappings $c:\IZ^d\to\IN$ assigning to any cell of the lattice $x\in\IZ^d$ the finite (non-negative) number $c(x)\in\IN$ of sand grains located in this cell.
The collection of all configurations is denoted as $\IN^{\IZ^d}$. Any configuration $c\in\IN^{\IZ^d}$ can also be considered as a \emph{macrostate}, or simply \emph{state}, of the sandpile model. Moreover the non-negative number $c(x)\in\IN$ of sand granules located by configuration $c$ at the cell $x\in\IZ^d$ of the lattice is the \emph{microstate} possesses by the involved cell. In this way, $\IN$ is the collection of all possible \emph{microstates} which can be assumed by any single cell.

Based on this general framework, the authors introduced:
\MP
(1) an invariant \emph{neighborhood} of any cell as a finite subset $\cN$ of $\IZ^d\setminus \{0^d\}$ (i.e., $\cN\incl \IZ^d$ s.t. $|\cN|<\infty$ and for any $x\in\cN$ it is $x\neq 0^d=(0,0,\ldots,0)$);
\\
(2) the \emph{distribution} of sand grains $\mathcal{D}:\cN\to\IN_+$ w.r.t. the neighborhood $\cN$ (let us stress that in any cell of the neighborhood $x\in\cN$ there must be located at least one granule).
\MP
A $d$--dimensional \emph{sandpile model} is so formalized by the triple $\para{d,\cN,\mathcal{D}}$ and on the basis of this notion it is defined the quantity $\vartheta :=\sum_{x\in\cN}\mathcal{D}(x)$, called the \emph{stability threshold}.

The \emph{discrete time dynamical system} based on the \emph{state space} of all configurations $\IN^{\IZ^d}$ is defined by a \emph{global transition function} $F:\IN^{\IZ^d}\to\IN^{\IZ^d}$ associating to any input configuration $c\in\IN^{\IZ^d}$ (i.e., mapping $c:\IZ^d\to\IN$) the output configuration $c'=F(c)\in\IN^{\IZ^d}$  (i.e., mapping $F(c):\IZ^d\to\IN$), generated in the formal context  $\para{d,\cN,\mathcal{D}}$ by the \emph{parallel} application to any cell $x$ of the \emph{local rule} defined by FP in \cite{FP20} through the formula:
$$
\oppA x\in\IZ^d,\; c'(x)=(F(c))(x):= c(x) -\vartheta \mathrm{H}(c(x)-\vartheta) + \sum_{y\in\cN} \mathcal{D}(y)\mathrm{H}(c(x+y)-\vartheta)
\leqno{(2)}
$$
(where $\mathrm{H}(r)=1$ if $r\in[0,+\infty)$, and $\mathrm{H}(r)=0$ otherwise, is the Heaviside function on the real domain $r\in\IR$).
In \cite{FP20} the authors claimed that $F$ is the global rule of the sandpile dynamics obtained by the \emph{parallel} application of the following \emph{local rule}: \virg{if a cell $x$ has at least $\vartheta$ grains, then it redistributed $\vartheta$ of its grain to its \emph{neighborhood} $x+\cN$ according to the distribution $\cD$.} This is a \virg{translation} on the supposed sandpile context of the following Goles statement about the \virg{chip firing game} \cite{Go92}: \virg{the application of the local rule consists of selecting a site [i.e., $x$] which has at least as many chips [i.e., FP grains] as its threshold $z_x$ [i.e., FP $\vartheta$] and passing one chip [i.e., FP grain] to each of its neighboring sites}.

In order to discuss from the foundational point of view the exact role of the local rule (2) as description of the parallel dynamics of some kind of \emph{generalized} sand pile (whatever the meaning to be attributed to the term \virg{generalized} -- but we will treat this in detail below), in the present paper we consider and deeply discuss the \emph{simplest} one--dimensional situation ($d=1$) characterized by the \emph{regular} neighborhood $\cN=\parg{-1,+1}$ and the \emph{constant} distribution function $\cD(x)=1$ for $x=\pm 1$, whose induced stability threshold is $\vartheta=2$. With respect to these choices the FP local rule (2) assumes the form
$$
\oppA x\in\IZ,\; c'(x)= c(x) -2 \mathrm{H}(c(x)-2) +\mathrm{H}(c(x-1) -2 ) +\mathrm{H}(c(x+1)-2)
\leqno{(2a)}
$$
This one-dimensional formulation will be the main argument of our investigation, organized according to the following parts:
\begin{description}
\item[First part]
In which we introduce and discuss the standard approach to the one-dimensional sandpile dynamics making reference to the seminal papers of Goles \cite{Go92} and Goles--Kiwi (GK) \cite{GK93}, these latter inspired by \cite{BTW87}, \cite{BTW88}.

First of all we realise that, differently from the GK approach in which the one--di\-men\-sio\-nal lattice of cells is $\IN$, their parallel sandpile dynamics can be naturally extended, without any formal difficulty, to the one-dimensional lattice of cells $\IZ$ proving in section \ref{sc:1sp-par} that this standard approach is governed by the global transition assigning to any initial configuration $c\in\IN^{\IZ}$ the next time configuration $c'\in\IN^{\IZ}$ expressed by the local rule:
$$
\oppA x\in\IZ,\; c'(x) = c(x) + \mathrm{H}(c(x-1) - c(x) -2) - \mathrm{H}(c(x)-c(x+1) -2).
\leqno{\text{(1a)}}
$$
This is the correct one--dimensional sandpile dynamics of the standard approach currently adopted by the scientific sandpile community. But from a comparison of equations (2a) and (1a) it is immediate to conclude that quantity $c(x)$ in the FP equation seems to have little to do with sandpile number of grains located in cell $x$ of the lattice of the standard GK approach (or at least that is our opinion).

On the other hand, if according to \cite{GK93} one introduces the \emph{height difference between consecutive piles} $h(x)= c(x) - c(x+1)$, in section \ref{sec:da-c-a-h} we prove that from equation (1a) one obtains the following rule:
$$
\oppA x\in\IZ,\; h'(x) = h(x) -2 \mathrm{H}(h(x)-2) +\mathrm{H}(h(x-1)-2) + \mathrm{H}(h(x+1)-2)
\leqno{\text{(2h)}}
$$
which has the same form of the supposed FP sandpile one--dimensional local rule (2a), but which in the standard approach to sandpiles more properly describes height differences between consecutive piles. From this point of view, the FP interpretation of equation (2a), or more generally of equation (2), is in contrast with the standard interpretation given to the same equations from the sandpile scientific community (see the above quoted papers).

Of course, as in Euclidean geometry triangles are triangles and circles are circles, without identifying circles with triangles, in the presently discussed case sandpile number of grains are sandpile number of grains and height differences are height differences, without any uncorrect identification of height differences with number of grains.
\item[Second part]
In this second part, despite the seen above correct GK interpretation of equation (2h) as describing height difference between consecutive piles, we want to explore the possible consequences of the totally different FP interpretation of equation (2a) as describing some kind of granules, of which we must identify the real identity (sand granules? Ice granules?), submitted to some kind of dynamics more complicated of the sandpiles one.

The conclusion we arrive in a formal way is that the correct interpretation of equation (2a) is of local rule of a \emph{spurious symmetric icepile parallel dynamics}, also obtained as suitable \emph{sequential} applications of the following three kind of rules:
\begin{enumerate}[(S{I}P1)]
\item
Vertical rules either from left to right $(VR)_d$ or its dual from right to left $(VR)_s$ typical of the symmetric sandpiles of \cite{FMP07}.
\item
Icepile horizontal rules, either of flowing from left to right $(HR)_d$ or its dual from right to left $(HR)_s$, in presence of horizontal plateaus.
\item
Bottom-up jump of an ice granule of one height either from left to right $(BT)_d$ or its dual from right to left $(BT)_s$.
\end{enumerate}
\end{description}
Rules (SIP1) and (SIP2) define the dynamics of symmetric \emph{icepiles}; it is the rule (SIP3) which assigns the \emph{spurious} dynamical behaviour to this model, which in any way cannot be considered a model of sandpiles but of something new kind of granules. 

%% file: spF-part1.tex
\section{Standard Goles-Kiwi (GK) one-dimensional sandpiles formal model under the vertical local rule}\label{sc:1sp-mpdel}
Let us start this section with a quotation of E. Formenti, B. Masson and T. Pisokas (FMP) from \cite{FMP07} where it is clear that in this paper the authors follow the usual standard definition of sandpile model according to the Goles-Kiwi (GK) approach: \virg{A formal model of sandpiles, called SPM, has been introduced in \cite{GK93, GMP02, GMP02a}. Each column contains a certain number of sand grains. The evolution is based on a local interaction rule: a sand grain falls from a column $A$ to its right neighbor $B$ if $A$ contains at least two granules more than $B$: otherwise there is no movement. The SPM hase been widely studied \cite{Br73, GK93, RS92, DRSV95, MN99, Mi99}.}

\begin{figure}[ht]
\begin{center}
   \includegraphics[width=3cm]{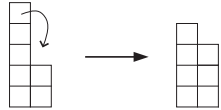}
\end{center}
      \caption{Typical local rule of the left--to--right granule movement of a sandpile}
      \label{fig:lr-gr-mov}
\end{figure}

Coherently with the just quoted FMP statement we introduce now the standard definition of the sandpile \emph{local rule} formally describing the configuration updating in the one--dimensional case.
\begin{enumerate}[(NG1)]
\item
The \emph{one--dimensional lattice of cells} is the set $\IN$ of all non-negative integer numbers.
\item
A \emph{configuration} is a mapping $c:\IN\to\IN$ assigning to every cell $x\in\IN$ the number of granules $c(x)\in\IN$ located in this cell, under the non-increasing condition $c(0)\ge c(1)\ge c(l-1)\neq 0$, with all the remaining $c(x)=0$ for $x\ge l$.
\\
The collection of these configurations will be denoted as $(\IN^\IN)_d$, with the subscript $d$ = decreasing.
\item
The update of the granule number of the generic pair of cells located in the places
$x, x+1\in\IN\times\IN$ is formalized by the so--called \emph{vertical local rule}:
\begin{multline*}
\text{(VR)}\qquad\text{If}\quad c(x)-c(x+1)\ge 2,\;\text{then we have the transition}
\\
(\ldots,c(x),c(x+1),\ldots)\to (\ldots, c(x)-1,c(x+1)+1,\ldots)\qquad
\end{multline*}
\end{enumerate}
Quoting from \cite{Go92}: \virg{The sandpile dynamics is defined from the introduction of a local rule which takes into account a critical threshold $\vartheta[=2]$. When the eight difference [$c(x)-c(x-1)$] becomes higher than $\vartheta$, one grain of sand tumbles to the lower level. The threshold represents the maximum slope permitted without provoking an avalanche.}

It is immediate to realize that this definition, formalized in its \virg{compact} form to the case of the one--dimensional lattice of cells $\IN$, at first glance seems to be difficult to generalize to the case of the lattice of cells $\IZ$, in which are taken in consideration configurations $c\in\IN^\IZ$ not necessarily satisfying the decreasing constraint.

With the aim of achieving this generalization, assuming a small variation of the  Goles and Kiwi
equation at pag. 324 sec.\til 1.2 of \cite{GK93}, let us now consider the following definition.
\begin{definition}\label{df:sp-loc-rule}
The one--dimensional \emph{sandpile local rule} is defined as the configuration transition $(\ldots, c(x),\ldots)\to (\ldots, c'(y),\ldots)$, with $x$ and $y$ cells of a one--dimensional lattice, realized by the law:
$$
c'(y) = \begin{cases}
               c(y)-\mathrm{H}(c(x)-c(x+1)-2)&\text{if}\; y=x
               \\
               c(y)+\mathrm{H}(c(x)-c(x+1)-2)&\text{if}\; y=x+1
               \\
               c(y) &\text{otherwise}
        \end{cases}
\leqno{\text{(VR-a)}}
$$
\end{definition}

Taking into account the formal behavior of the Heaviside function involved in this equation
\begin{equation}\label{eq:H-cond}
\mathrm{H}(c(x)-c(x+1)-2) = \begin{cases}
                   1 &\text{if}\; c(x)-c(x+1) \ge 2
                   \\
                   0 &\text{if}\; c(x)-c(x+1) \le 1
                   \end{cases}
\end{equation}
the (VR-a) con be re-formulated in the following way:

\begin{align*}
(2.1a)\qquad \text{if}\;& c(x)-c(x+1)\ge 2,\quad\text{then}\quad c'(x)= c(x)-1\;\text{and}\;c'(x+1) = c(x+1)+1,
\\
(2.1b)\qquad \text{if}\;& c(x)-c(x+1)\le 1,\quad\text{then}\quad c'(x) = c(x)\;\text{and}\;c'(x+1)  = c(x+1).
\end{align*}
In this way, the above definition {(VR-a)} can be equivalently formalized by the following two conditions:
\LP
{(VRa-1)}\quad If $c(x)-c(x+1)\ge 2$, then we have the transition
$$
(\ldots,c(x),c(x+1)\ldots) \to (\ldots, c(x)-1,c(x+1)+1,\ldots)
$$
{(VRa-2)}\quad If $c(x)-c(x+1)\le 1$, then we have the transition
$$
(\ldots,c(x),c(x+1)\ldots) \to (\ldots, c(x),c(x+1),\ldots)
$$
\begin{remark}
As to this result we can state the following.
\LP
-- The (VR-a) can be applied to the general case of a not necessarily non-increasing configuration. Indeed, thanks to the behavior of the Heaviside function in equation \eqref{eq:H-cond}, whose definition is also valid under the non-increasing condition $c(x)-c(x+1)\le 1$, and then a fortiori for $c(x)-c(x+1)\le 0$, the (VR-a) can be applied also to the particular case $c(x)\le c(x+1)$, that is without asking the configuration non-increasing.
\\
-- Another remark that underlines the importance of the (VR-a) with respect to the (VR) is that variables $x$ and $y$, which appear in the (VR-a), are not tied to range on $\IN$ (as in the (VR) case) but can range on $\IZ$, without \virg{conflicting} with the definition. In this way the (VR-a) can be applied to general configurations $c: \IZ\to \IN$, whose lattice of cells is the whole $\IZ$.
\end{remark}

The second point of this remark can be formulated in the following assumption.
\begin{itemize}
\item
{\it The sandpile local rule (VR-a) of definition \ref{df:sp-loc-rule} is applied to the collection $\IN^\IZ$ of all configurations defined on the whole one--dimensional lattice of cells $\IZ$, i.e., to all mappings $c:\IZ\to\IN$.}
\end{itemize}
The collection $\IN^\IZ$ of all configurations can be decomposed in the two following classes:
\begin{enumerate}
\item
A configuration $c\in\IN^\IZ$ is said to be \emph{stable} iff $\oppA x\in\IZ$, $c(x)-c(x+1)\le 1$;
\item
a configuration $c\in\IN^\IZ$ is \emph{unstable} iff $\oppE x_0\in\IZ$ s.t. $c(x_0)-c(x_0+1)\ge 2$.
\\
In the sequel we will say that the unstable configuration $c\in\IN^\IZ$ presents a \emph{critical jump} (or a \emph{critical slope}) in the pair of cells $x_0,x_0+1$.
\end{enumerate}
Moreover, in the development of the theory we are also interested to some particular subsets of configurations according to the following definitions.
\begin{itemize}
\item
{\it The collection $(\IN^\IZ)_f$ of all configurations $c:\IZ\to\IN$ of \emph{finite support}, i.e., such that \emph{[}$\oppE x_0\in\IZ$ s.t. $c(x_0)\neq 0$ and $c(x)=0$ for all $x< x_0$\emph{]} and \emph{[}$\oppE x_f\in\IZ$ s.t.\til $c(x_f)\neq 0$ and $c(x)=0$ for all $x > x_f$\emph{]}. In this case the support of the configuration $c$ is the finite subset of cells supp$(c):=\parg{x\in\IZ: x_0\le x \le x_f}$.
\item
The configuration $c:\IZ\to\IN$ is of \emph{(simply) connected support} iff it is of finite support and $\oppA x\in\text{supp}(c)$, $c(x)\neq 0$. Note that for any pair of points $a,b\in\text{supp}(c)$, with $a\le b$, the interval $\overline{a,b}:=\parg{z\in\IZ: a\le z\le b}$ is contained in $\text{supp}(c)$ (see the remark \ref{rk:simp-conn} below).
\item
Generalizing what seen above on the non-increasing configurations over $\IN$, we will also take into account $(\IN^\IZ)_d$ as the peculiar space of \emph{finite support} configurations with non-increasing values (let $x_0$ be the first element in supp$(c)$ such that $c(x_0)\neq 0$, then $c(x_0)\ge c(x_0+1)\ge \ldots \ge c(x_f)\neq 0$ and $c(x)=0$ for all $x> x_f$).
}
\end{itemize}
\begin{remark}\label{rk:simp-conn}
Let us recall the definition of \emph{simply connected subset} of $\IZ$. First of all, a \emph{bounded interval} of extreme points $a,b\in\IZ$, with $a\lvertneqq b$, is defined as $\overline{a,b}:=\parg{z\in\IZ: a\le z\le b}$.

Then, a subset $A\incl\IZ$ is \emph{simply connected} iff for any pair of points $a,b\in A$ the corresponding bounded interval $\overline{a,b}\incl A$. Examples of simply connected subsets of $\IZ$ are finite supports $\overline{x_0,x_f}$, not bounded intervals $\overline{x_0,\infty}:=\parg{z\in\IZ: x_0\le z}$ and $\overline{-\infty, x_0}:=\parg{z\in\IZ: z\le x_0}$. The subset $\overline{x_0,x_f}\cup \overline{y_0,y_f}$, with $x_f<y_0$, is not simply connected.
\end{remark}

To any configuration of finite support $c\in(\IN^\IZ)_f$ the sum $N(c):=\sum_{x\in\IZ} c(x)$ is finite and define the \emph{total number of granules} present in the configuration. The following result is trivial to prove.
\begin{lemma}
Let us consider the condition (VRa-1) of Definition \ref{df:sp-loc-rule} which, under the condition of critical jump on site $x\in\IZ$, $c(x)-c(x+1)\ge 2$, characterizes the configuration transition $c=(\ldots,c(x),c(x+1),\ldots) \in\IN^\IZ \to c'=(\ldots,c(x)-1,c(x+1)+1,\ldots)\in\IN^\IZ$.

If configuration $c$ is of finite support with total number of granules $N(c)$, then configuration $c'$ possesses the same finite support with the same total number of granules $N(c')=N(c)$.
\\
(The total number of granules is an \emph{invariant} quantity of the system).
\end{lemma}

Of course, a configuration $c\in\IN^\IZ$ has no finite support, i.e., it is of \emph{infinite support}, iff $\oppA x\in\IZ$, $\oppE x_0\le x$ s.t.\til $c(x_0)\neq 0$ and $\oppE x_1\ge x$ s.t.\til $c(x_1)\neq 0$; in this case the total number of granules is $N(c)=+\infty$.

Once clarified the context $\IN^\IZ$ of the configuration space in which to develop the theory,
as well known \virg{two dynamics can be defined [on the basis of a given local rule]: the sequential and the parallel update. The sequential one consists to update sites, one at time, in a prescribed order. For the parallel dynamics, all the sites are updated synchronously} \cite{Go92}.
As to this argument, let us also quote an adjustment to sandplies from \cite{GK93} originally related to \emph{chip firing game} (see the inserted square brackets):
\virg{The dynamics associated to the [(VR-a)] can be \emph{sequential} or \emph{parallel}. The sequential one consists in updating the [cells], one by one in a [...] prescribed periodic order. The parallel dynamics, which is the most usual one in the context of cellular automata, consists in updating all the [cells] synchronously.}
In the same \cite{GK93} paper, but in the specific section 1.2, titled \emph{the sandpile model}, we can quote \virg{The sandpile model simulates the avalanches produced in a one-dimensional profile of a sandpile. [...] The dynamics is specified as follows: a grain of sand tumbles from site $x$ to site $x+1$, iff the height difference $c(x)-c(x+1)$, is at least [...] 2. Clearly, 2 represents a critical slope of the sandpile. If the local slope of the sandpile at a specific site is at least 2, then an avalanche will occur at that site.}

Finally, from \cite{CCB12}: \virg{A discrete time dynamical system of the sandpiles is introduced by a vertical rule, also called (VR) rule, which solves any jump from left to right, greater that or equal to two granules. The (VR) rule can be applied in a sequential or in a parallel procedure. In the first case only one jump is solved step-by-step, whereas in the parallel case all the jumps are solved during a unique step by a synchronous application of the (VR) rule.}
\subsection{One-dimensional sequential sandpiles on the lattice of cells $\IZ$}
\label{sc:1sp-seq}
Let us anticipate that in the present first part of the paper  we have little interest in the sequential update procedure. But just wanting to take a quick look at this topic, we will examine the simple situation of configurations having finite support on the one-dimensional lattice $\IZ$ of cells. In this particular case, as seen in the two quoted papers of Goles \cite{Go92} and Goles-Kiwi \cite{GK93}, the sequential procedure consists in fixing a given order in the support of any configuration, for instance from left to right, and then update the involved microstates one at time according to this order.
\footnote{In the case of a configuration with non-finite support, a possible hypothetical order for the sequential updating of the cells can be the following: $0,1, -1,2, -2, \ldots$ and so on, but according to a theoretically infinite procedure.}

This procedure can be better explained with an example, where we adopt the convention of inserting the symbol $|$ in a configuration $c\in\IN^\IZ$ to denote that the integer value at its right corresponds to the cell of position $0$ in the lattice $\IZ$, in symbols $c=(\ldots, c(-1),|c(0),c(+1),\ldots)$.
\begin{example}\label{ex:5421}
Let us consider the finite support configuration
$c_0=(\bar{0},|5,4,2,1,\bar{0})$ as the initial state of a procedure, consisting in the sequential application of the sandpile local vertical rule (VR-a) from left to right.
Precisely, one can perform a finite sequence of \emph{levels} according to the following steps.
\begin{description}
\item[Level $L=0$] consisting of  the unique initial configuration $c_0$.
\item[Level $L=1$] consisting of the configuration $c_1=(\overline{0},|5,3,3,1,\overline{0})$  obtained by the application of the local vertical rule (VR-a) to the unique critical jump $4,2$ of the previous level $0$ configuration $c_0$.
\item[Level $L=2$] consisting of the two configurations $c_2^{(1)}=(\overline{0},|4,4,3,1,\overline{0})$ and $c_2^{(2)}= (\overline{0},|5,3,2,2,\overline{0})$, each obtained from the application of the local vertical rule (VR-a), the first on critical jump $5,3$ and the second on critical jump $3,1$.
\item[Level $L=3$] also this level consists of two configurations $c_3^{(1)}=(\overline{0},|4,4,2,2,\overline{0})$ and $c_3^{(2)}=(\overline{0},|5,3,2,1,1,\overline{0})$. While the second is the result of solving the single critical jump 2,0 of configuration $c_2^{(2)}$, the first is the result of solving the two critical jumps: 3,1 of configuration $c_2^{(1)}$ and 5,3 of configuration $c_2^{(2)}$.
\\ And so on. All other levels can be obtained straightforward without giving their formal construction.
\end{description}
The whole procedure as result of the just described levels can be represented by the following digraph of the configuration transitions $c_i\to c_{i+1}$ (or adopting the Brylawski notation of covering $c_i\succ c_{i+1}$ \cite{Br73}) from a level $L=i$ to its successive $L=i+1$, in which only the supports of the involved configurations are highlighted.

{\footnotesize{
\begin{figure}[h!]
$$\xymatrix{
{} & 5,4,2,1 \ar[d] &{}
\\
{} & 5,3,3,1 \ar[dl]\ar[dr] & {}
\\
4,4,3,1\ar[d] & {} & 5,3,2,2\ar[dll]\ar[d]
\\
4,4,2,2\ar[d]\ar[drr] & {} & 5,3,2,1,1\ar[d]
\\
4,3,3,2 \ar[dr] & {} & 4,4,2,1,1 \ar[dl]
\\
{} & 4,3,3,1,1 \ar[d]& {}
\\
{} & 4,3,2,2,1 & {} }
$$
\caption{}
\label{fg:5421}
\end{figure}
}}
\newpage

According to \cite{CCB12} the \emph{admissible} or \emph{possible paths} (also \emph{orbits, trajectories}) starting from the initial configuration $c_0$ are finite sequences of configurations depending from the time variable $t$,
$\gamma_{c_0}\equiv c_0 , c_1,\ldots, c_t,\ldots, c_{eq}$, constructed
according to the following points: (AP1) the initial configuration at time $t=0$ is $c_0$; (AP2) the configuration $c_{t+1}$ at time $t+1$ is obtained from the configuration $c_t$ at time $t$ by the application of the local vertical rule (VR-a) to a single critical jump inside it; (AP3) the final configuration is the equilibrium configuration $c_{eq}$. Of \emph{equilibrium} in the sense that it does not show any critical jump, and therefore it makes no sense to apply the local vertical rule to some of its cells: the procedure stops at this level of updating.

The following are all the admissible paths (orbits, trajectories) in the present example, each of initial configuration $c_0=(\overline{0},|5,4,2,1,\overline{0})$ and of final equilibrium configuration $c_{eq}=(\overline{0},|4,3,2,2,1,\overline{0})$; note that this equilibrium configuration has no critical jump inside it (all jumps are sub-critical $-4,1,0$).
\begin{align*}
\gamma_{c_0}^{(1)}\equiv\;&5,4,2,1\to 5,3,3,1\to 4,4,3,1\to 4,4,2,2\to 4,3,3,2\to 4,3,3,2,1\to 4,3,2,2,1
\\
\gamma_{c_0}^{(2)}\equiv\;&5,4,2,1\to 5,3,3,1\to 5,3,2,2\to 4,4,2,2\to 4,4,2,1,1\to 4,3,3,2,1\to 4,3,2,2,1
\\
\gamma_{c_0}^{(3)}\equiv\;&5,4,2,1\to 5,3,3,1\to 5,3,2,2\to 4,4,2,2\to 4,3,3,2\to 4,3,3,2,1\to 4,3,2,2,1
\\
\gamma_{c_0}^{(4)}\equiv\;&5,4,2,1\to 5,3,3,1\to 5,3,2,2\to 5,3,2,1,1\to 4,4,2,2,1\to 4,3,3,2,1\to 4,3,2,2,1
\end{align*}
\end{example}
\subsection{One-dimensional parallel sandpiles on the lattice of cells $\IZ$ as reformulation of the Goles-Kiwi approach on the lattice $\IN$}
\label{sc:1sp-par}
\par\noindent\\
Remaining in the context of the one-dimensional lattice of cells $\IZ$, whose configuration space is $\IN^\IZ$ (without any non-increasing requirement), in this section
we discuss the $\IN^\IZ$ sandpile model where the \emph{parallel} upgrading of the initial number of grains $c(x)\in\IN$ in the cell placed in $x\in\IZ$ towards the final number of grains
$c'(x)\in\IN$, always in the cell placed in $x$, is obtained by the following \emph{local rule} re-formulation on the lattice $\IZ$ of the equation (3) of \cite{GK93} formalized by the authors in the limiting context of the lattice of cells $\IN$.
\\
$\oppA c\in\IN^\IZ$ and $\oppA x\in\IZ$,
$$
c'(x) = c(x) + \mathrm{H}(c(x-1) - c(x) -2) - \mathrm{H}(c(x)-c(x+1) -2).
\leqno{\text{(1a)}}
$$

Let us summarize the theoretical context of application of local rule (1a) adopted in the present paper.
\begin{enumerate}[(GSPM1)]
\item
the \emph{lattice of cells} is the whole set of integer numbers $\IZ$;
\item
\emph{configurations} are mappings $c:\IZ\to\IN$ associating to any cell of the lattice $x\in\IZ$ the number $c(x)\in\IN$ of sand grains located in it. The collection of all configurations is then $\IN^\IZ$;
\item
the \emph{local rule} (1a) is applied to any configuration $c\in\IN^\IZ$ in a parallel way generating the \emph{global transition function} $F:\IN^\IZ\to\IN^\IZ$.
\end{enumerate}

In order to realize the main differences of this (GSPM) context with the Goles-Kiwe approach we now enumerate the main points of their approach.
\begin{enumerate}[(SPM1)]
\item
the \emph{lattice of cells} is the set of non negative integer numbers $\IN$;
\item
\emph{configurations} are mapping $c:\IN\to\IN$ associating to any cell of the lattice $x\in\IN$ the number $c(x)\in\IN$ of sand grains located in it under the \emph{non-increasing} condition. In this case the collection of all configuration is denoted as $(\IN^\IN)_d$;
\item
the \emph{local rule} (1a) is so applied to any configuration $c\in(\IN^\IN)_d$ in a parallel way generating the \emph{global transition function} $F:\IN^\IN\to\IN^\IN$.
\end{enumerate}

The local rule (1-a) responds to the main requests to describe the dynamics of a one-dimensional sandpile on $\IZ$ described in definition \ref{df:sp-loc-rule}. Indeed, we have the following cases depending from the behavior of the Heaviside involved values:
\begin{enumerate}[(SPZ1)]
\item
If the triple $c(x-1),c(x),c(x+1)$ is such that $$\mathrm{H}(c(x-1) - c(x) -2) = \mathrm{H}(c(x)-c(x+1) -2) = 1$$ then in equation (1-a) one has $c'(x)=c(x)$. The identities involving the Heaviside functions translate in the following two relationships, both of which are non-negative:
$c(x-1) - c(x) -2\ge 0$ e $c(x)-c(x+1) -2\ge 0$, which lead to the inequality chain
$$
c(x+1)+2\le c(x) \le c(x-1) -2
$$
In other terms, it must be $c(x-1)=c(x)+h$, with $h\ge 2$ and $c(x+1)=c(x)-k$, with $k\ge 2$, corresponding to a triple $c(x)+h,c(x),c(x)-k$, and under these conditions equation (1-a) furnishes the result $c'(x)=c(x)$. Formally,
for any $h,k\ge 2$
$$
c(x)+h,c(x),c(x)-k \;\longrightarrow\;  *\,,c(x),\,*
$$
where the symbol $*$ denotes an unknown value depending from the peculiar values assumed by $h$ and $k$.

From another point of view, with respect to the formulation (VR) of the vertical rule applied to the pair $c(x)+h,c(x)$ ($h\ge 2$) the central pile $c(x)$ gains a granule from the previous adjacent pile, but with respect to the pair $c(x),c(x)-k$ ($k\ge 2$) the same central pile $c(x)$ loses a granule towards the successive adjacent pile. As a final result the total number of grains of the cell $x$ stays invariant: $c'(x)=c(x)$.

\begin{figure}[h!]
\begin{center}
   \includegraphics[width=7cm]{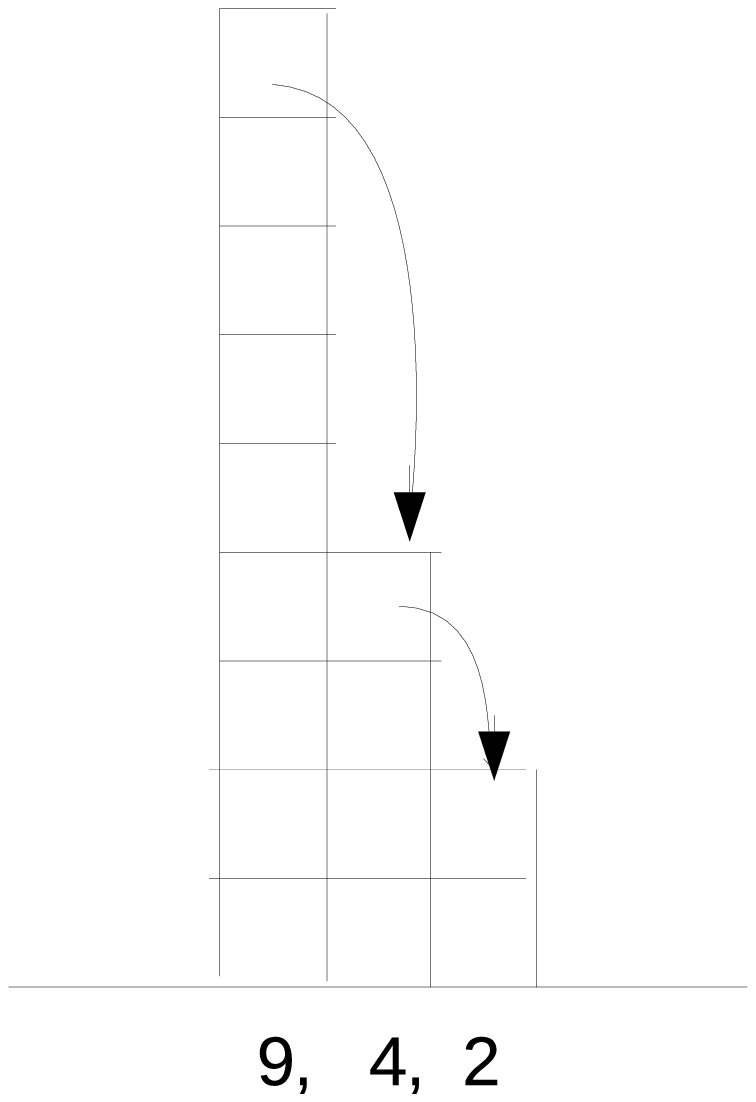}
\end{center}
      \caption{Triple $9,4,2$ as example of (SPZ1) situation. The local rule (VR) applied to the critical jump $9,4$ produces a gain of a granule to the central pile $4\to 4+1$, but the same rule applied to the critical jump $4,2$ produces a lost of a granule to the same pile $4\to 4-1$. The final result is that in the central pile there is no variation of the granule number, $9,4,2 \to *,4,*$ (the number $4$ of granules remains invariant).}
      \label{fig:SPZ1}
\end{figure}
\item
If the triple $c(x-1),c(x),c(x+1)$ is such that $$\mathrm{H}(c(x-1) - c(x) -2) = \mathrm{H}(c(x)-c(x+1) -2) =0$$ then in the equation (1-a) $c'(x)=c(x)$. But the identities involving the Heaviside functions translate in the two negative conditions:
$c(x-1) - c(x) -2 < 0$ and $c(x)-c(x+1) -2< 0$, or equivalently $c(x-1) - c(x) -1\le 0$ and $c(x)-c(x+1) -1\le 0$, which lead to the chain of inequalities
$$
c(x-1)-1\le c(x)\le c(x+1)+1
$$

\begin{figure}[h!]
\begin{center}
   \includegraphics[width=7cm]{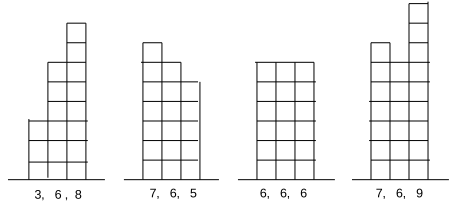}
\end{center}
\vspace{-1cm}
      \caption{Four examples of triples $c(x-1),c(x),c(x+1)$ satisfying the chain of inequalities $c(x-1)-1\le c(x) \le c(x+1)+1$ without any critical jump between the pairs of piles $x-1,x$ and $x,x+1$. The number of granules in the central cell remains invariant.}
      \label{fig:SPZ2}
\end{figure}
\newpage
\item
If the triple $c(x-1),c(x),c(x+1)$ is such that
$$\mathrm{H}(c(x-1) - c(x) -2)=0\quad\text{and}\quad \mathrm{H}(c(x)-c(x+1) -2) = 1$$
then the central cell $x$ loses a granule, $c'(x)=c(x)-1$, and this happens when
\\
$c(x-1)-c(x)-2<0$ e $c(x)-c(x+1)-2\ge 0$, i.e., when
$$c(x-1)-1\le c(x)\quad\text{and}\quad c(x+1)+2\le c(x).$$
Let us note that from the second inequality and from condition $c(x+1)\ge 0$ it follows that necessarily it must be $c(x)\ge 2$, from which we have that $c'(x)\ge 1$ (in any case strictly positive $c'(x)>0$).
\vspace{-2cm}
\begin{figure}[h!]
\begin{center}
   \includegraphics[width=7cm]{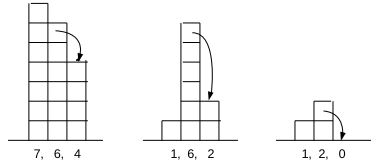}
\end{center}
\vspace{-1cm}
      \caption{Three examples of triples  $c(x-1),c(x),c(x+1)$ satisfying the conditions $c(x-1)-1\le c(x)$ and $c(x+1)+2\le c(x)$ with critical jumps at the pairs of cell $x,x+1$ corresponding to a loss of one granule in the central cell.}
      \label{fig:SPZ3}
\end{figure}
\item
If the triple $c(x-1),c(x),c(x+1)$ is such that
$$\mathrm{H}(c(x-1) - c(x) -2)=1\quad\text{and}\quad \mathrm{H}(c(x)-c(x+1) -2) = 0$$
then the central cell $x$ gains a granule, $c'(x)=c(x)+1$, and since $c(x)\ge 0$ we get that $c'(x)\ge 1$ (at any rate strictly positive $c'(x)>0$). This happens when
$c(x-1)-c(x)-2\ge 0$ and $c(x) -c(x+1)-2<0$, i.e., when
$$
c(x)\le c(x-1)-2\quad\text{and}\quad c(x)\le c(x+1)+1
$$
\vspace{-1cm}
\begin{figure}[h!]
\begin{center}
   \includegraphics[width=7cm]{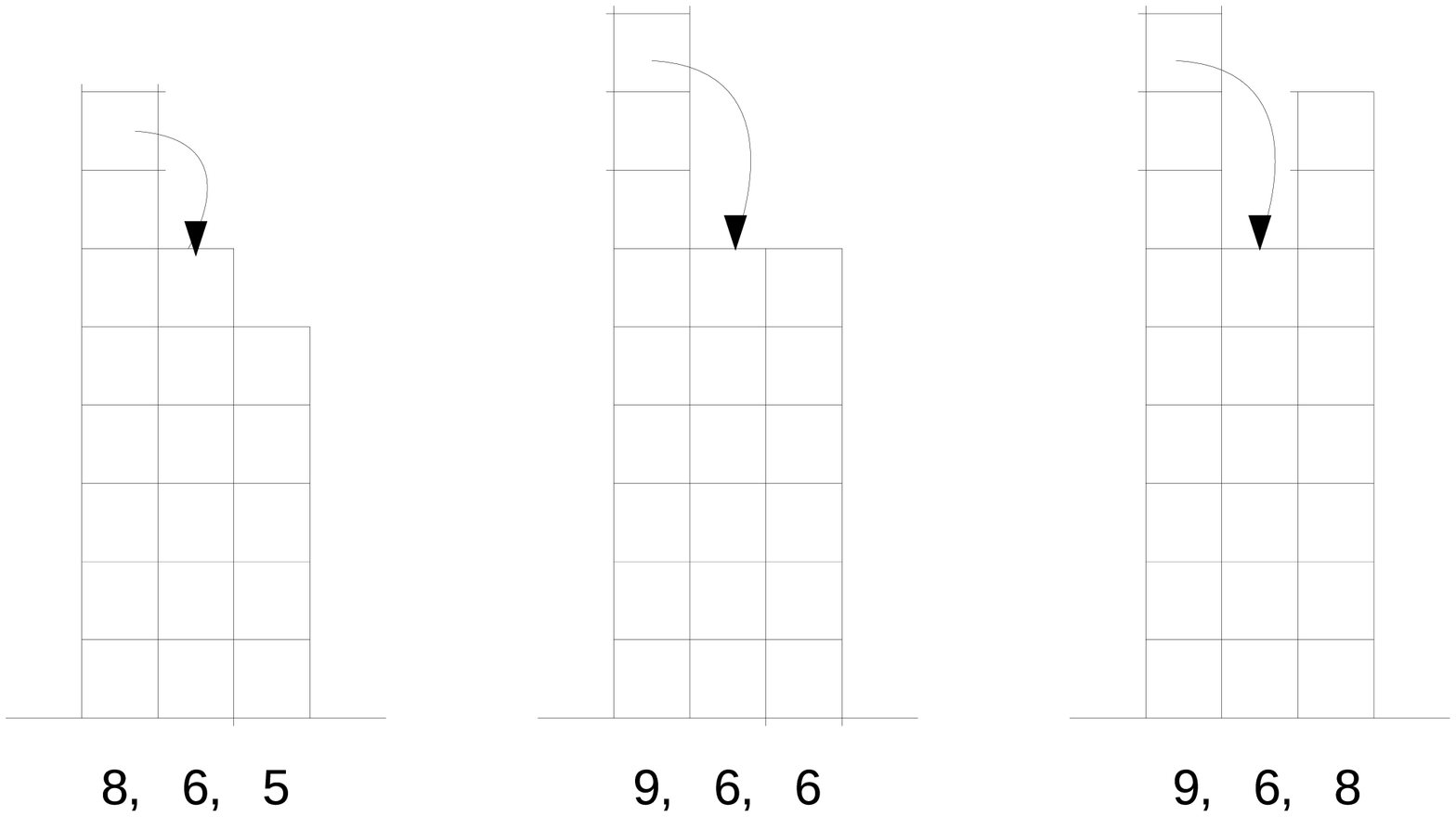}
\end{center}
\vspace{-1cm}
      \caption{Three examples of triples  $c(x-1),c(x),c(x+1)$ satisfying the conditions $c(x)\le c(x-1)-2$ and $c(x)\le c(x)+1$ with critical jumps at the pairs of cell $x-1,x$ corresponding to a gain of one granule in the central cell.}
      \label{fig:SPZ4}
\end{figure}
\newpage
\end{enumerate}

From the analysis of these four cases we see that condition  $c'(x)\ge 0$ holds for any $x\in\IZ$. This guarantees that the final configuration $c'\in\IN^\IZ$, in other words that the local rule (1a) generates a global transition $c\in\IN^\IZ\xrightarrow{F} c'\in\IN^\IZ$.
Moreover, owing to the fact that the Heaviside function can take only the two values 1 or 0, these four possibilities (i.e., 11,\,00,\,01,\,10) exhaust all the possible cases relatively to the pair of terms $\mathrm{H}(c(x-1) - c(x) -2)$ and $\mathrm{H}(c(x)-c(x+1) -2)$ appearing in equation (1a), thus providing all the possible local behaviors of this equation.

The following is an example of global dynamics generated by the local rule (1a), in its only four possibilities (SPZ1)--(SPZ4).
\begin{example}
The initial configuration $c_0=(\bar{0},|1,6,4,2,2,0,\bar{0})$, by the parallel application of the local rule (1a), is transformed in the updated configuration $c_1=F(c_0)=(\bar{0},|1,5,4,3,1,1,\bar{0})$.

The configuration $c_0$ presents three critical jumps $6,4$, $4,2$ and $2,0$, and the parallel global transition $c_0\xrightarrow{F}\;c_1$ is the result of the following single transitions of the involved triplets:
\begin{equation*}
\begin{tabular}{|c|c|}
\hline
\text{Transitions} & \text{Rules}
\\
\hline
$0,0,1\to *,0,*$ & \text{(SPZ2)}
\\
$0,1,6\to *,1,*$ & \text{(SPZ2)}
\\
$1,6,4\to *,5,*$ &\text{(SPZ3)}
\\
$6,4,2\to *,4,*$ &\text{(SPZ1)}
\\
$4,2,2\to *,3,*$ & \text{(SPZ4)}
\\
$2,2,0\to *,1,*$ & \text{(SPZ3)}
\\
$2,0,0\to *,1,*$ & \text{(SPZ4)}\\
\hline
\end{tabular}
\end{equation*}

Let us note that in the global transition $c_0\to c_1=F(c_0)$ the single granule movement happens, when it happens, from the left (where a granule is lost) towards the right (where a granule is gained).
For instance, the adjacent triple $6,4,2$ is transformed in the triple $5,4,3$ and the pair $2,0$ in the pair $1,1$.
All this coherently with the parallel application of the vertical rule (1a).
\end{example}
\subsubsection{The transition from the one-dimensional case on $\IZ$ to the case on $\IN$}
\label{ss:Z-to-N}
In the one-dimensional context of the lattice of cells $\IZ$, described by the local rule of equation (1a), if one considers the input configuration $c=(\bar{0},0,c(x_0)\neq 0,\ldots)$ then in the output configuration the cell of place $x_0-1$ is in the microstate $c'(x_0-1) = 0 + \mathrm{H}(0-0-2) - \mathrm{H}(0-c(x_0)-2)= 0$.
Moreover, for all the other cell of places $x_0-n$, $n\ge 2$, trivially the corresponding microstate is $c'(x_0-n)=0$.
In conclusion we obtain the parallel global transition $(\overline{0},0,c(x_0)\neq 0,\ldots)\xrightarrow{F}\;(\overline{0},0,c'(x_0),\ldots)$, whatever be the local transition of the microstate at place $x_0$, $c(x_0)\to c'(x_0)$.
\begin{description}
\item[Conclusion 1]
If on the lattice of cells $\IZ$ one has a configuration in which all the cells from $-\infty$ to $x_0-1$ have \emph{no} granules (i.e., they are empty) while $c(x_0)\neq 0$, then the local rule (1a) guarantees that during the whole parallel update all these cells remain empty. Formally, under the parallel action of the local rule (1a) one has the following global transition.
Let $c(x_0)\neq 0$, then
$$
(\bar{0},0,c(x_0),\ldots)\xrightarrow{\;F\;} (\bar{0},0,c'(x_0),\ldots)
$$
Moreover, if one takes into account that $\mathrm{H}(c(x_0-1)-c(x_0)-2)=\mathrm{H}(-c(x_0)-2)=0$, the parallel sandpile global dynamics on $\IZ$ according to the local rule (1a) for configurations of the type
$(\bar{0},0,c(x_0),c(x_0+1),\ldots)$, with $c(x_0)\neq 0$, is formalized in the following local rule behaviour  relative to the cell $x_0$:
$$
c'(x_0)= c(x_0) -\mathrm{H}(c(x_0)-c(x_0+1)-2)
$$
\end{description}

In other words, the parallel sandpile dynamics on $\IZ$ for configurations of the kind
\linebreak
$(\bar{0},0,c(x_0),c(x_0+1),\ldots)$ can be identified with the  parallel sandpile dynamics on the lattice of cells $\IN(x_0)=(x_0,x_0+1,\ldots,x_0+n,\ldots)$ for configurations $(c(x_0),c(x_0+1),\ldots)$ \virg{specified by the following local rule:
\begin{align*}
(1-i)\qquad&c'(x_0)=c(x_0)-\mathrm{H}(c(x_0)-c(x_0+1)-2)
\\
(1-ii)\qquad&c'(x)=c(x)+\mathrm{H}(c(x-1) -c(x) -2) -  \mathrm{H}(c(x) - c({x+1}) -2),\quad\oppA x>x_0.
\end{align*}
For this latter updating scheme we define the global transition function $F$ as
$$
F:\IN^{\IN(x_0)}\to\IN^{\IN(x_0)}, \quad c\to c'=F(c)
$$
where $\oppA x\in{\IN(x_0)}$, $c'(x)=(F(c))(x)$ is defined [according to the pair of equations] (1-i,1-ii).} (From Goles \& Kiwi \cite{GK93} in which these last considerations are treated in the particular case of $x_0=0$).
\subsection{The cellular automata interpretation of the parallel sandpile model and related deterministic dynamics}
\label{sec:GK-CA}
\par\noindent\\
Coming back to the general sandpile theory on the one dimensional lattice of cells $\IZ$, let us see as the sandpile local rule expressed by equation (1a) can be obtained in the context of a one-dimensional cellular automata (CA) model on the same lattice of cells.
Precisely, let us consider the one dimensional \emph{elementary} CA $\para{d,\mathcal{A},r,f}$ of \emph{dimension} $d=1$, based on the infinite \emph{alphabet} $\mathcal{A}=\IN$, of radius $r=1$ and \emph{local rule} given by the mapping $f:\IN^3\to\IN$, formally defined as follows:
\\
$\oppA (v,a,w)\in\IN^3$,
$$
f(v,a,w)= a + \mathrm{H}(v-a-2) - \mathrm{H}(a-w-2) =
\begin{cases}
a-1 &\text{if $v-1\le a$ and $w+2\le a$}
\\
a+1 &\text{if $a\le v-2$ and $a\le w+1$}
\\
0   &\text{otherwise}
\end{cases}
$$
Relatively to such elementary CA structure the \emph{discrete time dynamical system} (DTDS) is the pair $\para{\Omega,F_f}$ where the \emph{state space} is the collection $\Omega=\IN^\IZ$ of all bi-infinite sequences $c:\IZ\to \IN$ and the \emph{next state mapping} induced from the local rule is the mapping $F_f:\IN^\IZ\to\IN^\IZ$ transforming the input state $c\in\IN^\IZ$ into the output state $F_f(c)\in\IN^\IZ$  specified by the law
\\
$\oppA x\in\IZ$,
$$
[F_f(c)](x)=f\big(c(x-1),c(x),c(x+1)\big)
$$
Denoting  $c'(x):=[F_f(c)](x)$, this output CA state can also written as follows
\\
$\oppA x\in\IZ$,
$$
c'(x)=f(c(x-1),c(x),c(x+1))=c(x) + \mathrm{H}(c(x-1) - c(x) -2) - \mathrm{H}(c(x)-c(x+1) -2).
$$
which formally is just the final number of grains located in the cell $x$ expressed in subsection \ref{sc:1sp-par} by equation (1a) of the one-dimensional parallel approach to sandpile (SP): $\oppA x\in\IZ$, $[F_f(c)](x)=[F(c)](x)$ or, in other words, we can identify the two maps $F_f=F$.

For any fixed configuration $c_0$ this CA next state mapping, or equivalently SP global transition function, $F$ induces an
\emph{orbit} (or \emph{trajectory}, or also \emph{path}) of initial state $c_0$,
$$
\gamma_{c_0}\equiv c_0\xrightarrow{F} c_1=F(c_0)\xrightarrow{F} c_2=F(c_1)=F^2(c_0)\xrightarrow{F}\ldots
$$
described by the sequence of configurations
$$\gamma_{c_0}:\IN\to\IN^\IZ,\: t\to \gamma_{c_0}(t)=c_t,$$
where the general state of this dynamical evolution is expressed by the law
$$
\oppA t\in\IN\setminus\{0\},\quad c_{t}=F(c_{t-1}) = F^t(c_0).
$$
This orbit satisfies the following two Cauchy conditions of a first order difference equation:
$$
\begin{cases} \gamma_{c_0}(t+1) =F(\gamma_{c_0}(t))&\text{for every time instant}\; t\in\IN
\\
\gamma_{c_0}(0)=c_0
\end{cases}
$$

In this dynamical context the following definition turns out to be very important.
\begin{definition}
For definition $c_{eq}\in\IN^\IZ$ is an \emph{equilibrium configuration} iff $F(c_{eq})=c_{eq}$, since if at a time instant $\widehat t$ the dynamical evolution reach this state, $F^{\widehat t}(c_0)=c_{eq}$, then in any successive time instant $t\ge \widehat t$, it is $F^t(c_0)=c_{eq}$.
\end{definition}
\begin{lemma}\label{lm:eq-conf}
The configuration $c_{eq}\in\IN^\IZ$ is an \emph{equilibrium configuration} of the parallel dynamics generated by the local rule (1a) iff it is a stable configuration, i.e., $\oppA x\in\IZ$, $c_{eq}(x)-c_{eq}(x+1)\le 1$ (which can be defined as condition of \emph{sub-critical} jump).
\end{lemma}
\begin{proof}
The condition of equilibrium, for definition, is  $F(c_{eq})=c_{eq}$, and from equation (1a) this condition is equivalent to $\oppA x\in\IZ$, $\mathrm{H}(c_{eq}(x-1) - c_{eq}(x) -2) = \mathrm{H}(c_{eq}(x)-c_{eq}(x+1) -2)=0$. From the property of the Heaviside function, these two identities are equivalent to the two conditions
$\oppA x\in\IZ$, $c_{eq}(x)-c_{eq}(x+1)\le 1$ and $\oppA x\in\IZ$, $c_{eq}(x-1) - c_{eq}(x)\le 1$. But this second is nothing else that the first identity since putting in this latter $x=\widehat{x}+1$ we get $\oppA\widehat{x}\in\IZ$, $c_{eq}(\widehat{x})-c_{eq}(\widehat{x}+1)\le 1$.
\end{proof}
\begin{example}\label{ex:1354321}
Let us consider the configuration in $\IN^\IZ$ expressed by the sequence
$(\overline{0},1,3|5,4,3,2,1,\overline{0})$, depicted in the following figure.
\vspace{-2cm}
\begin{figure}[h!]
\begin{center}
   \includegraphics[width=7cm]{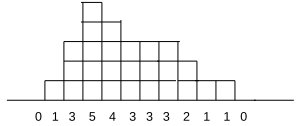}
\end{center}
      \caption{Example of a finite support equilibrium configuration, according to the previous Lemma \ref{lm:eq-conf}.}
      \label{fig:543}
\end{figure}

The following table shows the satisfaction of all the conditions $\oppA x\in\IZ$, $c(x)-c(x+1)\le 1$ expressed in Lemma \ref{lm:eq-conf} in order to have an equilibrium configuration.
\begin{equation*}
      \begin{tabular}{|c|c|c|}
         \hline
          $c(x)$ & $c(x+1)$ & $c(x)-c(x+1)$ \\
         \hline\hline
         \mbox{\rule[0cm]{0cm}{2.5ex} $0$} & 1 & $-1$ \\
         1 & 3 & $-2$\\
         $3$ & $5$ & $-2$ \\
         5 & 4& 1\\
         4 & 3 & 1\\
         3 & 3 & 0\\
         3 & 3 & 0\\
         3 & 2 & 1\\
         2 & 1 & 1\\
         1 & 1 & 0\\
         1 & 0 & 1\\
         \mbox{\rule[0cm]{0cm}{1.5ex} $0$} & 0 & 0\\
         \hline
      \end{tabular}
\end{equation*}
\end{example}
\begin{example}\label{ex:bool-equil}
The following is an interesting example of equilibrium configuration in which each cell contains at least a unique sand granule, i.e., it is a Boolean configuration $\oppA x\in\IZ$, $c(x)\in\parg{0,1}$.

\vspace{-4cm}
\begin{figure}[h!]
\begin{center}
   \includegraphics[width=8cm]{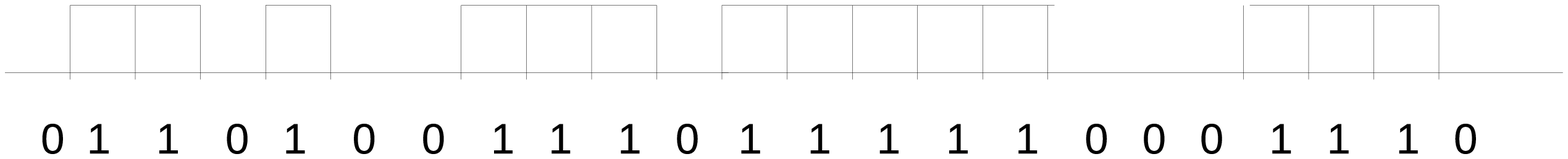}
\end{center}
\vspace{-1cm}
      \caption{}
      \label{fig:2-eq}
\end{figure}

Of course, any Boolean configuration is of equilibrium since it trivially presents sub-critical jumps $\oppA x\in\IZ$, $c(x)-c(x+1)\in\parg{-1,0,1}$.
\end{example}
\begin{example}
The configuration: $(\overline{0}|5,4,3,\overline{0})$ is not of equilibrium. Indeed, as shown by the following table not all the conditions required by Lemma \ref{lm:eq-conf} are satisfied.

\begin{equation*}
      \begin{tabular}{|c|c|c|}
         \hline
          $c(x)$ & $c(x+1)$ & $c(x)-c(x+1)$ \\
         \hline\hline
         \mbox{\rule[0cm]{0cm}{2.5ex} $5$} & 4 & $1$ \\
         $4$ & $3$ & $1$ \\
         3 & 0& 3\\
         \mbox{\rule[0cm]{0cm}{1.5ex} $0$} & 0 & 0\\
         \hline
      \end{tabular}
\end{equation*}

In particular, rule (1a) generates the dynamical transition $(\overline{0}|5,4,3,0,\overline{0})\xrightarrow{F} (\overline{0}|5,4,2,1,\overline{0})$, consequence of the presence of the unique critical jump $3,0$.
\end{example}
\begin{example}\label{ex:815}
Let us consider the configuration $c_0=(\overline{0},|8,1,5,\overline{0})$ as initial state of the dynamical evolution
$c _t \xrightarrow{F} c_{t+1}$
obtained by the parallel application of the local rule (1a). In the following table we represent this dynamical evolution, which ends after $t=6$ time steps with the equilibrium configuration $c_{eq}=(\overline{0},|4,4,3,2,1,\overline{0})$, exposing on the right the total number of granules of each configuration of the orbit.
\begin{align*}
c_0&=8,1,5 & N(c_0)&=14
\\
c_1&=7,2,4,1 & N(c_1)&=14
\\
c_2&=6,3,3,2 & N(c_2)&=14
\\
c_3&=5,4,3,1,1 & N(c_3)&=14
\\
c_4&=5,4,2,2,1 & N(c_4)&=14
\\
c_5&=5,3,3,2,1 & N(c_5)&=14
\\
c_6&=4,4,3,2,1=c_{eq} & N(c_6)&=14
\end{align*}
Of course, $\oppA t\ge 6$, $c_t=c_{eq}= (\overline{0},|4,4,3,2,1,\overline{0})$.
Moreover, this orbit  satisfies the principle of \emph{invariance of the total number of granules}
$
\oppA t\in\IN,\; N({c_t}) =\sum_{x\in\IZ} c_t(x)=14.
$
\end{example}
The final result about the invariance of the total number of grains during the dynamical evolution is not an \virg{accident} of this particular example. Indeed the following general result can be proved.
\begin{proposition}\label{pr:N-invariance}
For any initial state $c_0$ the corresponding orbit $c:\IN\to\IN^\IZ$, $t\to c_t=F^t(c_0)$ satisfies the principle of \emph{invariance of the total number of granules}
$$
\oppA t\in\IN,\; N(c_t) =\sum_{x\in\IZ} c_t(x)= \sum_{x\in\IZ} c_0(x) = N(c_0).
$$
\end{proposition}
\begin{example}\label{ex:5421-N}
Below we will refer to the example \ref{ex:5421} discussed in the subsection \ref{sc:1sp-seq}. In particular in the left column of the figure \ref{fg:5421-bis} we will treat the sequential procedure with dashed lines, while we will draw the parallel procedure with continuous lines. In the central column of the same figure we isolate the only case of parallel dynamics while in the column on the right we highlight the invariance behavior of the total number of granules $N=12$ for both sequential and parallel dynamics.
{\scriptsize{
\begin{figure}[h!]
$$\xymatrix{
{} & 5,4,2,1 \ar[d] &{}
&{}& 5,4,2,1 \ar[d] &{} & N=12
\\
{} & 5,3,3,1 \ar@{-->}[dl]\ar@{-->}[dr] \ar[ddl]& {}
&{}& 5,3,3,1 \ar[dd] &{} & N=12
\\
4,4,3,1\ar@{-->}[d] & {} & 5,3,2,2\ar@{-->}[dll]\ar@{-->}[d]
&{}& {} &{} & N=12
\\
4,4,2,2\ar@{-->}[d]\ar@{-->}[drr] \ar[ddr]& {} & 5,3,2,1,1\ar@{-->}[d]
&{}& 4,4,2,2 \ar[dd] &{} & N=12
\\
4,3,3,2 \ar@{-->}[dr] & {} & 4,4,2,1,1 \ar@{-->}[dl]
&{}& &{} & N=12
\\
{} & 4,3,3,1,1 \ar[d]& {}
&{}& 4,3,3,1,1\ar[d] &{} & N=12
\\
{} & 4,3,2,2,1 & {}
&{}& 4,3,2,2,1 &{} & N=12
}
$$
\caption{}
\label{fg:5421-bis}
\end{figure}
}}
\newpage

Of course, $(\overline{0},|4,3,2,2,1,\overline{0})$ is the equilibrium configuration of the parallel dynamics since all the conditions expressed by Lemma \ref{lm:eq-conf} of sub-critical jumps are satisfied.
\end{example}

%% file: spF-part2.tex
\subsection{Sandpile dynamics of initial state whose support is perfect (finite and simply connected)}
\par\noindent\\
Let us observe that in both examples \ref{ex:815} and \ref{ex:5421-N} treated in subsection \ref{sec:GK-CA} the dynamics involved converge to an equilibrium configuration in a finite number of steps, keeping the total number of granules constant.

In this section we want to demonstrate that this behavior is not an exceptional fact but represents two particular cases of a general behavior of sandpiles dynamical evolution whose initial configuration $c_0\in\IN^\IZ$ is of perfect (i.e., finite and simply connected) support and of total number of granules $N$.

Before addressing this topic, let us introduce an interesting result, where for simplicity we denote by $[0,N-1]=\parg{0,1,\ldots,N-1}$ (resp., $[0,N]=\parg{0,1,2,\ldots,N}$).
\begin{lemma}\label{lm:N!}
The collection $[0,N]^{[0,N-1]}$ is of finite cardinality equal to $N!$. Formally,
$$
|[0,N]^{[0,N-1]}|=N!
$$
\end{lemma}
\begin{proof}
In order to prove this relationship we adopt the procedure using in statistical thermodynamics.

We have to count the number of mappings $\parg{0,1,2,\ldots,N-1}\to\parg{0,1,2,\ldots,N}$. Let us consider the co-domain $\parg{0,1,2,\ldots,N}$ as consisting of $N+1$ originally empty boxes, and the domain $\parg{0,1,2,\ldots,N-1}$ as consisting of $N$ distinguishable balls. Fixing the first box $y=0$ it can be filled by a ball in $N$ different ways.
At this point there are $N-1$ balls with which the second box $y=1$ can be filled.
Thus, the pair of boxes $1,2$ can be filled by balls in $N (N-1)$ different ways, leaving $N-2$ balls available.
Then, the third box $y=2$ can be filled in $N-2$ different ways and so the triple of boxes $1,2,3$  can be filled in $N(N-1)(N-2)$ different ways.

Continuing in this way until the balls are exhausted in filling all the available boxes, we arrive at the desired relationship.
\end{proof}

In the sequel let us denote by $\Omega(N)$ the \emph{state space}  collection of \emph{all} bi-infinite configurations $c=(\overline{0},|c(0),\ldots,c(l),\overline{0})\in\IN^\IZ$ such that:
\\
(1) they have $N$ as total number of granules ($\sum c(x)=N$);
\\
(2) according to subsection \ref{ss:Z-to-N} they can be considered as elements of $\IN^\IN$ in the sense that
$c(x)=0$ for every $x< 0$;
\\
(3) they are of perfect (i.e., finite and simply connected) support ($\oppE l\in\IN\setminus\{0\}$ s.t. $c(x)\neq 0$ for every $0\le x\le l$ and $c(x)=0$ for every $x>l$);
\\
(4) of length at most equal to $N$ ($l\le N$).
\begin{remark}
In a generic configuration $c=(\overline{0},|c(0),c(1),\ldots,c(x),\ldots, c(l),\overline{0})\in\Omega(N)$, the condition $c(0)+c(1)+\ldots+c(x)+\ldots+c(l)=N$, with $c(x)\ge 0$ for every $x$, defines $c$ as a \emph{generalized partition} of $N$.
So from this point of view $\Omega(N)$ is the collection of all \emph{generalized partitions} of $N$.

According to \cite{Br73} a \emph{partition} of $N$ (also \emph{ordered partition} according to \cite{GK93}) is a generalized partition satisfying the further condition of non-increasing: $c(0)\ge c(1)\ge \ldots \ge c(x)\ge \ldots \ge c(l)$.
The collection of all ordered partition of $N$ will be denoted by $S(N)$. So $S(N)\incl \Omega(N)$.
\end{remark}

Let us start this general investigation from the case in which the initial configuration $c_0\in\Omega(N)$, i.e., it is of the kind $c_0=(\overline{0},|c_0(0),\ldots,c_0(l_0),\overline{0})$, with $\sum_{k=0}^{l_0}c_0(k)=N$ and $\oppA 0\le j\le l_0$, $c_0(j)\neq 0$, underlining some important properties that characterize the dynamics.
\begin{enumerate}[(Pr1)]
\item
The initial configuration $c_0\in\Omega(N)$ will \virg{vary} between the two extremal cases
\linebreak
$0:=(\overline{0},|\underbrace{(1,1,\ldots,1)}_{N-times},\overline{0})$
and
$1:=(\overline{0},|N,\overline{0})$.
\item
From the condition $\sum_{k=0}^{l_0}c_0(k)=N$ it follows that $\oppA k$, $c_0(k)\le N$.
\item
From Proposition \ref{pr:N-invariance}, which assures the invariance of the total number of granules during the dynamical evolution, it follows that any state of the orbit $\gamma_{c_0}(t)=c_t$, for time $t\in\IN$, is of the kind $c_t=(\overline{0},|c_t(0),\ldots, c_t(l_t),\overline{0})$, with $0<l_t\le N$ and $\sum_{x=0}^{l_t}c_t(x)=N$. So any state $\gamma_{c_0}(t)=c_t\in\Omega(N)$.
\end{enumerate}
The previous properties allow us to consider $[0,N]^{[0,N-1]}$ as the framework of the state space $\Omega(N)$, and so also of $S(N)$, instead of $\IN^\IZ$.
\\
Summarizing, we consider the following three sets:
\begin{equation*}
\begin{tabular}{ll}
$[0,N]^{[0,N-1]}$&collection of mappings $c$ from $[0,N-1]$ to $[0,N]$,
\\
$\Omega(N)$       &collection of mappings $c$ from $[0,N-1]$ to $[0,N]$, s.t.\til $\sum_x c(x)=N$,
\\
$S(N)$            &collection of mappings $c$ from $[0,N-1]$ to $[0,N]$, s.t.\til $\sum_x c(x)=N$, non-increasing.
\end{tabular}
\end{equation*}
From the trivial chain of set inclusions  $S(N)\incl \Omega(N)\incl [0,N]^{[0,N-1]} $ and the fact that the cardinality of $[0,N]^{[0,N-1]}$ is finite and equal to $N!$ (Lemma \ref{lm:N!}), we have that also $S(N)$ and $\Omega(N)$ have finite cardinality with
\begin{equation}\label{eq:ub-SO}
|S(N)|\le |\Omega(N)| \le N!
\end{equation}

\section{The dynamical evolution from a non-increasing sandpile initial configuration}
\par\noindent\\
Let us summarize in the present subsection the dynamical evolution starting from a non-increasing initial state
as described in section 2, entitled \virg{\emph{General result about sandpiles}}, of \cite{GK93}.
To be precise, let us consider the collection of all the \emph{ordered partitions} of $N$
$$
S(N)=\Big\{c\in\IN^\IN:\oppA x\in\IN,\;c(x)\ge c(x+1),\,\sum_{x\in\IN} c(x)=N\Big\}\incl\Omega(N)
$$
Let us recall the two sandpile dynamics introduced in section \ref{sc:1sp-mpdel}:
\begin{enumerate}
\item[(S)]
The \emph{sequential} dynamics updates the cells of any configuration, one by one, in a prescribed order, in general from left to right, according to the \emph{vertical local rule} (VR).
\item[(P)]
The \emph{parallel} dynamics updates all the cells of any configuration synchronously, according to the \emph{local rule} (1a).
\end{enumerate}
Then the following is proved.
\\ \\
\textbf{Corollary 2.4 of \cite{GK93}}.\label{cor:GK93}
Given any initial configuration [$c_0\in S(N)$], both the SPM sequential and parallel dynamic converge towards the same fixed point, i.e., equilibrium state.
\par\noindent
[We can add the information that both the sequential and the parallel \emph{transient time} to reach the equilibrium configuration cannot be greater than the upper bond $N!$ of the cardinality $|S(N)|$ of the space $S(N)$ (see equation \eqref{eq:ub-SO}).]

Furthermore, the following properties are verified (from \cite{Go92}, \cite{GK93}).
\begin{enumerate}[(SN1)]
\item
For any non-negative integer $N$ there exist two non-negative integers $k,k'\in\IN$
such that it can be written as
$$
N=\frac{1}{2}\,k(k+1)+k',\; \text{with}\; k'\le k
$$
\item
Any sequential orbit (trajectory) of initial sandpile state concentrate in the origin $c_0=(N,\overline{0})\in S(N)$ converges to the fixed point (equilibrium state)
$$
c^{(k,k')}_{eq}=(k,k-1,\ldots,k-j=k'+1,k',k',k'-1,\ldots,2,1,\overline{0}),\quad \text{for}\; j=(k-k')-1
$$
\item
The sequential \emph{transient time} to reach the fixed point $c^{(k,k')}_{eq}$ is exactly
$$
T(N)=\binom{k+1}{3}+k\,k'-\binom{k'}{2}
$$
\end{enumerate}
\begin{example}
In the following figure we draw at the left side the sequential (VR) dynamical evolution and at the right side the  parallel one (PT), both of initial state $(6,0,0,0,0,0)$, of the sandpile model.
As expected from the general discussion, both dynamics converge in a finite number of steps to the same fixed point (equilibrium configuration) $(3,2,1,0,0,0)$.
\begin{figure}[h!]
$$
\xymatrix{ & & 6 & &
 6
\\
& & 5,1\ar@{<-}[u]^{(VR)} & &
5,1\ar@{<-}[u]_{(PT)}
\\
& & 4,2\ar@{<-}[u]^{(VR)} & &
4,2\ar@{<-}[u]_{(PT)}
\\
& 3,3 \ar@{<-}[ur]^{(VR)} & & 4,1,1 \ar@{<-}[ul]_{(VR)} &
{}
\\
& & 3,2,1 \ar@{<-}[ul]^{(VR)}\ar@{<-}[ur]_{(VR)} & &
3,2,1\ar@{<-}[uu]_{(PT)}
}
$$
\caption{}
\label{fig:060}
\end{figure}

In agreement with the above point (SN3), in the sequential dynamics the equilibrium state is reached after four time steps: indeed, from $6=\frac{3\cdot 4}{2}$ ($k=3$ and $k'=0$), we get $T(6)=\binom{4}{3} +3\cdot 0 +\binom{0}{2}= 4$.
The parallel dynamics reaches the same equilibrium state after three time steps.

The sequential dynamics can be considered as decomposition of the two orbits of the same initial state $c_0=(6,0,0,0,0,0)$, both converging to the same final equilibrium state $c_{eq}=(3,2,1,0,0,0)$:
\begin{align*}
\gamma_{c_0}^{(1)}=&6 \xrightarrow{(VR)}\; 5,1 \xrightarrow{(VR)}\; 4,2 \xrightarrow{(VR)}\; 3,3 \xrightarrow{(VR)}\; 3,2,1
\\
\gamma_{c_0}^{(2)}=&6 \xrightarrow{(VR)}\; 5,1 \xrightarrow{(VR)}\; 4,2 \xrightarrow{(VR)}\; 4,1,1 \xrightarrow{(VR)}\; 3,2,1
\end{align*}
\end{example}

%% file: spF-part3.tex
\subsection{Failure of the one-dimensional generalization of the local rule (1a) to the case of a generic neighborhood}
\label{ss:failure}
\par\noindent

Let us now write the local rule (1a) for the parallel upgrade of a one-dimensional sandpile on the lattice $\IZ$ in a form which can lead to a \emph{possible} generalization to the case of any neighborhood not containing the cell 0.
\begin{align*}
c'(x) &= c(x) + \mathrm{H}(c(x-1) - c(x) -2) - \mathrm{H}(c(x)-c(x+1) -2)
\\
      &= c(x) + \sum_{y\in\{-1,+1\}} y\, \mathrm{H}(y\,(c(x-y) - c(x))-2)
\end{align*}

From this formulation it follows that a possible one-dimensional generalization consists in introducing a \emph{neighborhood} $\mathcal{N}$ as a finite subset of $\IZ\setminus\{0\}$ and a \emph{generalized distribution function} $\mathcal{G}:\mathcal{N}\to \IN$ w.r.t. the neighborhood $\mathcal{N}$, with associated \emph{stability threshold} $\theta=\sum_{y\in\mathcal{N}} |\mathcal{G}(y)|$. Note that this distribution function is \emph{generalized} in the sense that it can assume not only positive values but also negative.
The \emph{global transition function} of a generalized one-dimensional sandpile model on $\IZ$ \emph{could} therefore be formalized as a mapping $F$ assigning to any input configuration $c\in\IN^\IZ$ the output configuration $F(c)\in\IN^\IZ$ obtained by the parallel application to any cell $x\in\IZ$ of the following \emph{local rule}:
\\
$\oppA c\in\IN^\IZ$, $\oppA x\in\IZ$,
$$
(F(c))(x):= c(x) + \sum_{y\in\mathcal{N}} \mathcal{G}(y)\, \mathrm{H}(\mathcal{G}(y)\, (c(x-y)-c(x))-\theta)
\leqno{\text{(1g)}}
$$

The sandpile global rule (1a) seen above is the particular case of this generalized global rule (1g) under the choices of $\mathcal{N}=\{-1,+1\}$ and $\mathcal{G}(y)=y$ for any $y=\pm 1$, from which it follows that $\theta=2$.

We introduced the generalized local rule (1g) using the conditional \virg{could} since all this makes sense if we prove the following
\begin{description}
\item[Open Question]
Since a configuration of a sandpile must associate to each cell the number, greater than or equal to zero, of granules allocated in this cell, the configuration $F(c)$ must be a quantity greater than or equal to zero in each cell of the lattice $x\in\IZ$, and this is a property that must be proved to be satisfied by (1g) for any cell of the lattice. Formally,
given the local rule (1g), the following non-negativity condition must be demonstrated:
\\
$\oppA\mathcal{N}\incl \IZ\setminus\{0\}$ with $|\mathcal{N}|<\infty$ and $\oppA\mathcal{G}\in{\IN}^\mathcal{N}$, let $c\in\IN^\IZ$ then
$$
\oppA x\in\IZ,\quad (F(c))(x)\in\IN.
\leqno{\text{(NN)}}
$$
\end{description}
This condition is problematic to prove given the great arbitrariness in the choice of the distribution function $\mathcal{G}:\mathcal{N}\to \IN$. For instance, given a generic neighborhood $\mathcal{N}$ a possibility is the following distribution function $\oppA y\in\mathcal{N}$, $\mathcal{G}(y)=-\text{exp}(\sqrt[3]{\arctan(y)})$.
\\
In a first approach we could consider the two quite simple cases $\oppA y\in\mathcal{N}$, $\mathcal{G}_{id}(y)=y$ and $\mathcal{G}_1(y)=1$, with respect to which the first result is the following.
\begin{lemma}
If the neighborhood $\mathcal{N}$ in $\IZ$ is \emph{symmetric}, i.e., $y\in\mathcal{N}$ implies $\,-y\in\mathcal{N}$, then the condition of non-negativity (NN) is verified for the constant generalized distribution $\mathcal{G}_{1}$, but in general the conservation of the total number of granules is not verified.
\end{lemma}
\begin{proof}
Under the symmetry condition of $\mathcal{N}$, in the sum of (1g) the following pairs of terms appear
$$
\mathrm{H}((c(x-y)-c(x))-\theta) + \mathrm{H}((c(x+y)-c(x))-\theta)
$$
whose contribution to the sum is one of the non-negative values 0, 1 and 2, which under the condition $c(x)\ge 0$ maintain  the non-negativity of (1g).

Let us see now a simple example where the number of granules conservation is not verified. Let us consider the neighborhood $\mathcal{N}=\{-1,+1\}$ and the constant generalized distribution on $\mathcal{N}$ equal to 1 with corresponding $\theta=2$. In this case, the local rule (1g) assumes the form:
$$
c'(x)=c(x) + \mathrm{H}(c(x-1)-2) + \mathrm{H}(c(x+1)-2)
$$
Given the configuration $c=(\bar{0},0,4,|0,4,0,\bar{0})$, whose number of granules is 8, then $c'=(\bar{0},1,4,|2,4,1,\bar{0})$ whose number of granules is 12.
\end{proof}
\begin{proposition}\label{pr:-y+y}
In the case of a neighborhood $\mathcal{N}=\{-y,+y\}$, with the integer number $y>0$ fixed, and of the identity generalized  distribution $\mathcal{G}_{id}(y)=y$ and $\mathcal{G}_{id}(-y)=-y$, the positivity condition (NN) is satisfied in the only two cases $y=1$ and $y=2$.
\end{proposition}
\begin{proof}
Under the hypothesis of the proposition, with respect to which $\theta=2y$, we get
\begin{align*}
c'(x)&=c(x) + y \mathrm{H}(y(c(x-y)-c(x))-2y) -y \mathrm{H}(-y(c(x+y)-c(x)) -2y)
\\
&=c(x) + y \mathrm{H}(y(c(x-y)-c(x))-2y) -y \mathrm{H}(y(c(x)-c(x+y)) -2y)
\end{align*}
Let us discuss all the possible cases with respect to the behaviour of the Heaviside function, whose possible values are 0 or 1.
\\
- If both the values of the two Heaviside functions are equal to zero ($0,0$) or equal to one ($1,1$) we have $c'(x)=c(x)\ge 0$, and so the positivity condition is verified.
\\
- If the first Heaviside function is equal to zero and the second equal to 1 (case 0,1) we have $c'(x)=c(x)+y\ge 0$ since $y>0$. So also in this case there is no problem with respect to the positivity.
\\
Let us note that this case corresponds to $y(c(x-y)-c(x))-2y\ge  0$ and $y(c(x)-c(x+y))-2y< 0$, i.e., when $c(x)\le c(x-y) -2$ and $c(x)<c(x+y)+2$, and this is a situation similar to the (SPZ4) when we put $x-y$ and $x+y$ in place of $x-1$ and  $x+1$.
\\
- If the first Heaviside function is equal to zero and the second equal to one (case 0,1) we have that $c'(x)=c(x)-y$.
This happens when the Heaviside arguments are $c(x-y)-c(x)-2<0$ and $c(x)-c(x+y)-2 \ge 0$, respectively, that is
$$
c(x-y)-1\le c(x)\quad\text{and}\quad 2\le c(x+y)+2\le c(x)
$$
The case $y=1$ corresponds to the previously discussed point(SPZ3), which showed no problem with respect to the non-negativity of $c'(x)$.
Also the case $y=2$ does not present any problem with respect to the non-negativity of $c'(x)$. Indeed, in this case  $c'(x)=c(x)-2$, with $c(x)\ge 2$.

All the cases $y\ge 3$ are problematic. Let us see only two cases, all the others are obtained accordingly.
\\
Let $y=3$ and $c(x)=2$, then $c'(x)=-1$.
\\
Let $y=4$. If $c(x)=2$, then $c'(x) =-2$; but if $c(x)=3$, then $c'(x)=-1$.
\end{proof}
\begin{description}
\item[Conclusion 2]
The generalization (1g) of the local rule (1a), although at first sight interesting, cannot be taken into consideration given the fact that already in cases of very simple symmetrical neighborhoods the condition of coherence (NN), which requires the non-negativity of the transformed $F(c)$ of each generic configuration $c$, is not satisfied.
This in order to be able to interpret the quantity $(F(c))(x)$ as the (non-negative) number of sand grains allocated in cell $x$ of the transformed configuration.
\end{description}
Let us note that the local rule (1g), in order to recover the two \virg{canonical} behaviors of the proposition \ref{pr:-y+y} for $y = 1$ and $y = 2$, requires a distribution function $\mathcal{G}_{id}$ which assumes negative values (i.e. it assumes values in $\IZ$ instead of the usual set $N_+$). This difficulty can be overcome by considering the following local rule:
\\
$\oppA c\in\IN^\IZ$, $\oppA x\in\IZ$,
$$
(F(c))(x):= c(x) + \sum_{y\in\mathcal{N}} \mathcal{D}(y)\,y\, \mathrm{H}(\mathcal{D}(y)\,y\, (c(x-y)-c(x))-\theta)
\leqno{\text{(1g')}}
$$
Of course, the canonical local rule (1a) can be obtained from this generalization (1g') for the neighborhood $\mathcal{N}=\{-1,+1\}$ and the distribution function $\mathcal{D}_1(-1)=\mathcal{D}_1(+1)=1$. But also in this case one can apply the proof of proposition \ref{pr:-y+y} in order to obtain the same result on the satisfaction of the non-negativity condition (NN) for only the two cases $y =1 $ and $y = 2$.
\section{From the one-dimensional dynamics on the number of granules $c(x)$, to the dynamics of height difference $h(x)$}
\label{sec:da-c-a-h}

In \cite{GK93} and \cite{Go92}, besides the parallel sandpile dynamics of the number of granules expressed by the local transition $x\to c(x)$ formalized by equation (1a), reference is also made to the derived dynamics of the height difference between successive positions along the sandpile:
$$
h(x):=c(x)-c(x+1)
\leqno{\text{(ED)}}
$$
where, for the moment, we have deliberately not specified the domain of variability of the $x$ position in the lattice of cells, which can be $\IN$ or $\IZ$.
\subsection{The case when the space is the lattice $\IZ$}
\par\noindent

Given a configuration $c:\IZ\to\IN$ the local rule performs the following two changes: 
\begin{align*}
c'(x) &= c(x)+\mathrm{H}(c(x-1) -c(x) -2) -  \mathrm{H}(c(x) - c({x+1}) -2)
\\
c'(x+1) &= c(x+1)+\mathrm{H}(c(x) -c(x+1) -2) -  \mathrm{H}(c(x+1) - c({x+2}) -2)
\end{align*}
Let $h(x) := c(x+1) - c(x)$ be the \emph{difference of heights}, relative to some initial configuration $c$, of two adjacent cells $x$ and $x+1$. We can infer that the difference of heights $h'(x)$ in configuration $c'$ is:
\begin{align*}
h'(x):=c'(x)-c'(x+1) = &[c(x)-c(x+1)] -2 \mathrm{H}([c(x)-c(x+1)]-2)
\\
&+\mathrm{H}([c(x-1)-c(x)]-2) + \mathrm{H}([c(x+1) - c({x+2})] -2)
\end{align*}
From the definition (ED), we can infer the \emph{local rule} for the parallel update of the difference of heights $\oppA x\in\IZ$:
$$
(\Phi(h))(x)=h'(x) := h(x) -2 \mathrm{H}(h(x)-2) +\mathrm{H}(h(x-1)-2) + \mathrm{H}(h(x+1)-2)
\leqno{\text{(2h)}}
$$
Notice that we moved from a rule updating the number of ``granules'' to a rule updating the difference in heights of consecutive cells. We can now analyse some important properties of this new local rule:
\begin{enumerate}[(D1)]
  \item
  The application of rule (2h) can result in negative values. For example with:
  $$
  c(x-1)=c(x)=0\qquad c(x+1) = 6,\; c(x+2)\ge 0
  $$
  we will have:
  $$
  c'(x)-c'(x+1) = -7 + \mathrm{H}(4-c(x+2)) =\begin{cases} -7 &\text{if}\; 5\le c(x+2)
    \\
    -6 &\text{if}\; c(x+2)\le 4
  \end{cases}
  $$
  In both cases this difference has a negative value. If we interpret this value as a height difference, according to (ED), then the sequence of differences $\{h'(x)=c'(x)-c'(x+1):x\in\IZ\}$ can assume negative values for some positions in $\IZ$.

  \item
  We can consider only configurations over $\IZ$ with finite support, like $(\bar{0},0,c(x_0),c(x_0+1),\ldots,c(x_0+l-1),\bar{0})$, under the following decrease condition: $c(x_0)\ge c(x_0+1)\ge\ldots\ge c(x_0+l-1)> 0$. \\
  Under these conditions, the configuration composed of the difference of heights assumes the form, $\oppA x\in\IZ$:
  $$
  h(x)=c(x)-c(x+1)=\begin{cases}
    0&\text{for}\; x< x_0-1
    \\
    -c(x_0) &\text{for}\; x =x_0-1
    \\
    c(x_0+n)-c(x_0+n+1) &\text{for}\; 0\le n < l-1
    \\
    c(x_0+l-1) &\text{for}\; x=x_0+l-1
    \\
    0 &\text{for}\; x>x_0+l-1
  \end{cases}
  $$
  thus resulting in:
  \begin{gather*}
    \oppA x\in\IZ\setminus\{x_0-1\},\quad h(x)\ge 0
    \\
    h(x_0-1)= -c(x_0)\le 0.
  \end{gather*}

  Under the condition that $c(x_0) \ge 2$ and from (2h), the value of $h'$ in the cell $x_0 - 1$ is:
  $$
  h'(x_0-1) = h(x_0-1)  -2 \mathrm{H}(h(x_0-1)-2) +\mathrm{H}(h(x_0-2)-2) + \mathrm{H}(h(x_0)-2)
  $$
  that is,
  $$
  h'(x_0-1) = -c(x_0) +\mathrm{H}(c(x_0)-c(x_{0}+1)-2)\le 0
  $$
\end{enumerate}

We can observe the behaviour we just described using an example from~\cite{Go92}, applied to the lattice $\IZ$ instead of the lattice $\IN$ as originally employed by Goles:

\begin{example}
  Let $\IZ$ be the lattice used in this example. The initial configuration at time $t = 0$ of the ``number of granules'' is $c_0=(\bar{0},0,|6,0,\bar{0})$, and the corresponding difference of heights is $h_0=(\bar{0},-6,|6,0,\bar{0})$. Therefore, the value $-6$ in the cell in position $-1$ is due to $h_0(-1)=c_0(-1)-c_0(0)=-6$.

  Let us see the first step in the dynamics of both configurations.

  \begin{description}
    \item[$t=1$] Since at time $t = 0$ between the cells in position $0$ and $1$ there is a critical jump $c_0(0) - c_0(1) = 6 > 2$, the local vertical rule (VR) is applied and at time $t = 1$ we will have that $c_1(0)=c_0(0)-1=5$ e $c_1(1)=c_0(1)+1=1$. That is, given that $\oppA x \in \IZ$, $h_1(x)=c_1(x)-c_1(x+1)$, we have:
    $$
    c_1=(\bar{0},0,|5,1,\bar{0})\quad\text{and}\quad h_1=(\bar{0},-5,|4,1,\bar{0})
    $$
    Notice how the difference in height of the cell $x = -1$ is negative: $h_1(-1)=c_1(-1) - c_1(0)=-5$.\\
    Let us notice that we obtained these results via the ``direct'' definition (VR) and (ED) of $c_1(x)$ and $d_1(x)$, with a results identical to the one obtained via the ``indirect'' formulae (1a) and (2a).
  \end{description}

  If we continue following the same cell update rules for all successive time steps, it is possible to obtain the following dynamics, summarised in the following table, which is substantially identical to table (iii) of Fig. 3 of \cite{Go92}:
  
  \begin{align*}
    {}&{} && \quad c_t && \;\quad h_t
    \\
    t& =0
          &&\bar{0},0|6
                       &&\bar{0},-6|6
    \\
    t&=1
          &&\bar{0},0|5,1
                       &&\bar{0},-5|4,1
    \\
    t&=2 && \bar{0},0|4,2 && \bar{0},-4|2,2
    \\
    t&=3 && \bar{0},0|3,2,1 && \bar{0},-3|1,1,1
  \end{align*}

  Clearly, the two configurations $c_3=(\bar{0},0,|3,2,1,\bar{0})$ and $h_3=(\bar{0},-3,|1,1,1,\bar{0})$ are fixed points, ($\oppA t>3$, $c_t=c_3$ and $h_t=h_3$) for the number of granules and the difference in heights, respectively.
\end{example}
\section{A generalization from the lattice of cells $\IZ$ to its $d$-dimensional version $\IZ^d$, but with the different interpretation of number of chips}
\par\noindent

Let us continue with formula (2h), which expresses in a functional way and in one dimensional case of the lattice $\IZ$ the ``local'' dynamics of the difference in heights $\oppA x\in\IZ$, $h(x)=c(x)-c(x+1)$ of the number of granules in cells $x$ and $x+1$. By looking at the neighbourhood $\mathcal{N}=\{-1,+1\}$ not containing the cell $0 \in \IZ$, the local dynamics given by (2h) can be reformulated as follows:
$$
\oppA x\in\IZ,\;\;(\Phi(h))(x) = h(x) - 2 \mathrm{H}(h(x)-2) +\sum_{y\in\mathcal{N}} \mathrm{H}(h(x+y)-2)
\leqno{(\text{2h}')}
$$

If we want to generalize to the $d$-dimensional case, i.e., the lattice $\IZ^d$, it is possible to generalize the notion of configuration as \emph{number of granules} in each cell $\oppA x=(x_1,x_2,\ldots,x_d)\in\IZ^d$, $c(x_1,x_2,\ldots,x_d)\in\IN$ (i.e., $c\in\IN^{\IZ^d}$).
\\
However, it would be difficult to generalize (1a) to the $d$-dimensional case since, at a first glance, it would be difficult to imagine, for a generic finite neighbourhood $\mathcal{N}$ in $\IZ^d$ without the cell $(0, 0, \ldots, 0)$, a generalized version for $c(x_1, x_2, \ldots, x_d)$ of the quantities $c(x-1)$ and $c(x+1)$.
\\
Hence, it would also be difficult to generalize to the $d$-dimensional case the one-dimensional notion of difference in heights $\oppA x\in\IZ$, $d(x):=c(x)-c(x+1)$.
\subsection{The $d$--dimensional semantics of Goles (and of Goles-Kiwi) chip firing game and of Formenti-Perrot granules}
However, on the other hand, as do Goles in \cite{Go92}, and Goles-Kiwi in \cite{GK93}, in the $d$-dimensional context of the lattice of cells $\IZ^d$ we must drop the semantic of height difference associated with a generic mapping $h:\IZ^d\to\IN$ (i.e., $h\in\IN^{\IZ^d}$), interpreting instead the quantity $h(x)\in\IN$ as the \emph{number of chips} located in the cell $x\in\IZ^d$ of a chip firing game.

Introduced this semantics, which therefore has nothing to do with the height difference between two neighboring columns of granules of the one-dimensional case, once fixed a finite neighborhood $\mathcal{N}$ not containing the origin $\vec{0}=(0,0,\ldots,0)$, whose cardinality will be denoted by $\theta=|\mathcal{N}|$, a generalization of the parallel dynamics induced from the local rule (2h') can be given as follows:
$$
\oppA x\in\IZ^d,\quad h'(x)=h(x) -\theta \mathrm{H}(h(x)-\theta) + \sum_{y\in\mathcal{N}}\mathrm{H}(h(x+y)-\theta)
\leqno{(2g)}
$$
Trivially, the local rule (2g) is a particular case of the Goles parallel dynamics specified by the local rule (1.2) in \cite{Go92}, rewritten below with respect to our notations,
$$
\oppA x\in\IZ^d,\quad h'(x)= h(x) -z(x) \mathrm{H}(h(x)-z(x)) + \sum_{r\in V(x)}\mathrm{H}(h(r) -z(r))
\leqno{(1.2)}
$$
Indeed, equation (2g) is obtained from Goles' equation (1.2) once fixed in this latter all the thresholds $z(x)=z(r)=\theta$, for every $x$ and every $r=x+y$, and assuming that all the neighbourhoods $V(x)$ are unchanged as $x$ varies, in such a way that $r\in V(x)$ (i.e., $x+y\in V(x)$) iff $y\in\mathcal N$.

Let us stress that for the dimension $d\ge 2$ we have two different semantical interpretations.
\begin{enumerate}[(S{I}1)]
\item
The Goles (and also Goles-Kiwi) \emph{firing game model} in which the non-negative integer $h(x)$ describes the number of chips located at the site $x$ and the equation (1.2) formalizes the local rule specifying the parallel dynamics.
\\
Quoting from \cite{Go92}: \virg{A site such that $h(x)\ge z(x)$ [i.e., $h(x) \ge\theta$ in our notation] will be called a firing site. [...] Equation (1.2) is interpreted as follows: a site $x$ loses $z(x)$ chips if its number of chips is at least $z(x)$ and receives one chip from each firing neighborhood}.
\\
As seen above, the (2g) is a particular case of the (1.2).
\item
In the multidimensional context of the lattice of cells $\IZ^d$ under the constant distribution function $\mathcal{D}(y) = 1$ for $y \in \mathcal{N}$, equation (2g) is indeed formally identical to the local rule (2) of Formenti--Perrot (FP) paper \cite{FP20} except for the different semantic interpretation of the configuration $h\in\IN^{\IZ^d}$, which in Goles describes the number of chips located in the cell $x\in\IZ^d$, whereas FP interpret it as the number of granules located in the same cell.
\end{enumerate}
In the one-dimensional case of $d=1$ the (2h') is a particular case for the neighborhood $\mathcal{N}=\{-1,+1\}$, with associated $\theta=2$. The interpretation (SI1) of chip firing game can obviously be maintained also in this particular one-dimensional case. We will now give the significant semantic interpretations of this particular one-dimensional case.
\begin{enumerate}[(fD1)]
\item
As widely discussed in section \ref{sec:da-c-a-h}, in the formulation of the parallel dynamics described by local rule (2h) the non-negative quantity $h(x)\in\IN$ is interpreted as the height difference of the number of sand grains located between the cells $x$ and $x+1$ of the one-dimensional lattice $\IZ$.
\item
The interpretation of $h(x)\in\IN$ as number of grains located in the site $x$ can be translated to the present very particular one-dimensional case, i.e., $x\in\IZ$, even if it is absolutely not correct to assign it the meaning of number of sand grains, as we will demonstrate in the next sections, but rather as the number of \emph{ice grains} of a particular bilateral model.
\end{enumerate} 

%% file: spF-part4.tex
\section{The improper Formenti--Perrot (FP) interpretation of number of chips $h$ as number of sand granules $c$ and related theory}
\label{sc:FP-inter}
As pointed out in the previous semantical interpretation (SI2), in equation (2) of the FP paper \cite{FP20} it is introduced a \emph{global transition function} $F:\IN^{\IZ^d}\to\IN^{\IZ^d}$ which applied to configurations  $c\in\IN^{\IZ^d}$, semantically interpreted as mapping associating to any cell of the lattice $x\in\IZ^d$ the \emph{number of sand grains} $c(x)\in\IN$, produces in a parallel update the successive configuration $F(c)\in\IN^{\IZ^d}$. Formally, such a global transition is defined by the following \emph{local rule} version:
\\
$\oppA x\in\IZ^d$, (i.e., $x=(x_1,x_2,\ldots, x_d)$),
$$
(F(c))(x)= c(x) -\vartheta \mathrm{H}(c(x)-\vartheta) + \sum_{y\in\mathcal{N}}\mathrm{H}(c(x+y)-\vartheta)
\leqno{(2-\text{FP})}
$$
where, as explained in the Introduction, $\mathcal{N}$ is a suitable finite neighborhood not containing the origin ($0\notin \mathcal{N}$) and $\theta=|\mathcal{N}|$ is the cardinality of this neighborhood called \emph{threshold}.

As done in the failed case of equation (1g), subsection \ref{ss:failure}, the delicate point is to prove the non-negative condition (NN), but in this regard the following result holds.
\begin{proposition}\label{pr:NN-d}
The local rule (2-FP) is such that, whatever be the configuration  $c\in\IN^{\IZ^d}$ and whatever be the \emph{finite} neighborhood $\mathcal{N}\incl\IN\setminus\{0\}$, the condition of non-negativity is satisfied, i.e.,
$$
\oppA x\in\IZ^d,\quad (F(c))(x)\in\IN.
$$
\end{proposition}
\begin{proof}
From the behavior of the Heaviside function of assuming the only two values, either 0 or 1, one gets that the sum in equation (2-FP)  always provides a non-negative contribution.
Therefore it remains to be analyzed the problematic part
$$
c_1(x):= c(x) -\vartheta \mathrm{H}(c(x)-\vartheta)
$$
The behavior of $c_1(x)$ depends on the behavior of the Heaviside function that appears in it. There are only two possible cases:
\\
(1) Case $c(x)-\vartheta < 0$. Since in this case $\mathrm{H}(c(x)-\vartheta)=0$, we get that $c_1(x)=c(x) \ge 0$, which does not cause problems on the non-negative sign of $(F (c))(x)$.
\\
(2) Case $c(x)-\vartheta \ge 0$. Since in this case $\mathrm{H}(c(x)-\vartheta)=1$, we get that $c_1(x)=c(x)-\vartheta$, which is non-negative by hypothesis.
So also this case does not involve problems on the non-negative sign of $(F(c))(x)$.
\end{proof}

To tell the truth in \cite{FP20} it is formalized a further generalization of this local rule, once introduced in addition to the neighborhood $\mathcal{N}$ a \emph{distribution function} $\mathcal{D}:\mathcal{N} \to \IN_+$ and its \emph{stability threshold} $\vartheta=\sum_{y\in\mathcal{N}}\mathcal{D}(y)$. This local rule is expressed by the law
\\
$\oppA x\in\IZ$, (i.e., $x=(x_1,x_2,\ldots, x_d)$),
$$
(F(c))(x)= c'(x)= c(x) -\vartheta \mathrm{H}(c(x)-\vartheta) + \sum_{y\in\mathcal{N}} \mathcal{D}(y) \mathrm{H}(c(x+y)-\vartheta)
\leqno{(2g-\text{FP})}
$$
Oviously, the proof of the non-negativity condition (NN) made in Proposition \ref{pr:NN-d} can be immediately extended to this case. Besides, in the particular case of the distribution function defined by the law $\oppA y\in\mathcal{N}$, $\mathcal{D}(y)=1$, we have that $\vartheta=|\mathcal{N}|$ and consequently (2g-FP) reduces to (2-FP).
\subsection{The one-dimensional case on the lattice $\IZ$ of the Formenti-Perrot model}
\par\noindent\\
Let us now consider the local rule (2-FP) in the one-dimensional case on the lattice $\IZ$, corresponding to the particular case of $\mathcal{N}=\{-1,+1\}$, with respect to which $\theta=2$, formally given by the law:
\\
$\oppA x\in\IZ$,
$$
(F(c))(x)= c'(x)= c(x) -2 \mathrm{H}(c(x)-2) + \mathrm{H}(c(x-1)-2) + \mathrm{H}(c(x+1)-2)
\leqno{(2a-\text{FP})}
$$

This local rule generates 8 cases of transitions $c(x-1),c(x),c(x+1)\to *,c'(x),*$ corresponding to the two possible values $0,1$ assumed by the Heaviside functions involved in its formal expression, which we divide in two groups depending on the behaviour of $c(x)$.  
\\ \\
\framebox{\textbf{Group: $c(x)\in\parg{0,1}$}}
\begin{enumerate}[(SFP1)]
\item
If $\mathrm{H}(c(x)-2)=\mathrm{H}(c(x-1)-2)=\mathrm{H}(c(x+1)-2)=0$, i.e., $c(x-1)\in\parg{0,1}$, $c(x)\in\parg{0,1}$, $c(x+1)\in\parg{0,1}$, then it happens the transition
$$c(x-1),c(x),c(x+1)\to *,c(x),*\quad\text{with}\;c(x)\in\parg{0,1}$$
\item
If $\mathrm{H}(c(x)-2)=\mathrm{H}(c(x-1)-2)=0$ and $\mathrm{H}(c(x+1)-2)=1$, i.e., $c(x-1)\in\parg{0,1}$, $c(x)\in\parg{0,1}$ and $2\le c(x+1)$, then it happens the transition
$$c(x-1),c(x),c(x+1)\to *,c(x)+1,*\quad\text{with}\;c(x)\in\parg{0,1}$$
\item
If $\mathrm{H}(c(x)-2)=0$, $\mathrm{H}(c(x-1)-2)=1$, $\mathrm{H}(c(x+1)-2)=0$, i.e., $2\le c(x-1)$, $c(x)\in\parg{0,1}$, $c(x+1)\in\parg{0,1}$, then it happens the transition
$$c(x-1),c(x),c(x+1)\to *,c(x)+1,*\quad\text{with}\;c(x)\in\parg{0,1}$$
\item
If $\mathrm{H}(c(x)-2)=0$ and $\mathrm{H}(c(x-1)-2)=\mathrm{H}(c(x+1)-2)=1$, i.e., $2\le c(x-1)$, $c(x)\in\parg{0,1}$, $2\le c(x+1)$, then it happens the transition
$$c(x-1),c(x),c(x+1)\to *,c(x)+2,*\quad\text{with}\;c(x)\in\parg{0,1}$$
\end{enumerate}
\framebox{\textbf{Group: $2\le c(x)$}}
\begin{enumerate}
\item[(SFP5)]
If $\mathrm{H}(c(x)-2)=1$ and $\mathrm{H}(c(x-1)-2)=\mathrm{H}(c(x+1)-2)=0$, i.e., $c(x-1)\in\parg{0,1}$, $2\le c(x)$, $c(x+1)\in\parg{0,1}$, then it happens the transition
$$c(x-1),c(x),c(x+1)\to *,c(x)-2,*\quad\text{with}\;2\le c(x)$$
\item[(SFP6)]
If $\mathrm{H}(c(x)-2)=1$, $\mathrm{H}(c(x-1)-2)=0$, $\mathrm{H}(c(x+1)-2)=1$, i.e., $c(x-1)\in\parg{0,1}$, $2\le c(x)$, $2\le c(x+1)$, then it happens the transition
$$c(x-1),c(x),c(x+1)\to *,c(x)-1,*\quad\text{with}\;2\le c(x)$$
\item[(SFP7)]
If $\mathrm{H}(c(x)-2)=1$, $\mathrm{H}(c(x-1)-2)=1$, $\mathrm{H}(c(x+1)-2)=0$, i.e., $2\le c(x-1)$, $2\le c(x)$, $c(x+1)\in\parg{0,1}$, then it happens the transition
$$c(x-1),c(x),c(x+1)\to *,c(x)-1,*\quad\text{with}\;2\le c(x)$$
\item[(SFP8)]
If $\mathrm{H}(c(x)-2)=\mathrm{H}(c(x-1)-2)=\mathrm{H}(c(x+1)-2)=1$, i.e., $2\le c(x-1)$, $2\le c(x)$, $2\le c(x+1)$, then it happens the transition
$$c(x-1),c(x),c(x+1)\to *,c(x),*\quad\text{with}\;2\le c(x)$$
\end{enumerate}

Also in the present case of FP parallel dynamics generated by the local rule (2a-FP) (similarly to the GK parallel dynamics of the standard sandpile model treated in subsection \ref{sec:GK-CA} -- on the other hand this is a general behavior of any discrete time dynamical system) we can introduce the notion of \emph{equilibrium configuration} as any configuration $c_{eq}\in\IN^\IZ$ such that $F(c_{eq})=c_{eq}$, i.e., it is a \emph{fixed point} of the global transition function $F:\IN^\IZ\to\IN^\IZ$.

The following result characterizes the form of configurations which are of equilibrium in the FP parallel dynamics.
\begin{lemma}\label{lm:FP-eq}
The configuration $c_{eq}\in\IN^\IZ$ is of \emph{equilibrium} with respect to the parallel dynamics governed by the local rule (2a-FP), i.e., $F(c_{eq})=c_{eq}$ iff $\oppA x\in\IZ$, $c_{eq}(x)\in\parg{0,1}$, i.e., it is a Boolean sequence.
\end{lemma}
\begin{proof}
Under the condition  $F(c_{eq})=c_{eq}$ the equation (2a-FP) leads to the condition $\oppA x\in\IZ$, $2\mathrm{H}(c_{eq}(x)-2) +\mathrm{H}(c_{eq}(x-1)-2) +\mathrm{H}(c_{eq}(x+1) -2)=0$, from which we have in particular that $\oppA x\in\IZ$, $\mathrm{H}(c_{eq}(x)-2) =0$, i.e., that $\oppA x\in\IZ$, $c_{eq}(x)\in\parg{0,1}$.

Under the condition $\oppA x\in\IZ$, $c_{eq}(x)\in\parg{0,1}$ the triplet $c_{eq}(x-1),c_{eq}(x),c_{eq}(x+1)\in\parg{0,1}^3$, i.e., it is a Boolean triplet, and so the corresponding transition (SFP1) is necessary of the type $c_{eq}(x-1),c_{eq}(x),c_{eq}(x+1)\to *,c_{eq}'(x)=c_{eq}(x),*$, that is $\oppA x\in\IZ$, $c'(x)=c(x)$; but since $c'(x)$ is a simplified formulation of $(F(c))(x)$ we have proved that $\oppA x\in\IZ$,  $(F(c))(x)=c(x)$.
\end{proof}
\begin{remark}
This result about the Boolean equilibrium configurations of the parallel (FP) model must be compared with the equilibrium configurations of the standard parallel (GK) sandpile model discussed in Lemma \ref{lm:eq-conf}, this latter characterized by the condition
\linebreak
$\oppA x\in\IZ$, $c_{eq}(x)-c_{eq}(x+1)\le 1$. Recall that in example \ref{ex:bool-equil} it is shown that all the Boolean configurations are of (GK) equilibrium.
\end{remark}
\begin{example}\label{ex:FP-eq}
The following are examples of FP equilibrium configurations of the parallel dynamics $(\bar{0},|0,1,1,0,0,0,0,1,0,1,\bar{0})$, $(\overline{0,1},|0,1,\overline{0,1})$, $(\overline{1},0,|1,0,\overline{1})$.
\end{example}

Making reference to the local rule (2g-FP) acting on configurations from the $d$-dimensional state space $c\in\IN^{\IZ^d}$, \virg{a cell $x\in\IZ^d$ is said to be \emph{stable} if $c(x)<\vartheta$ and \emph{unstable} otherwise. A configuration is \emph{stable} when all cell are stable, and is \emph{ustable} if at least one cell is unstable. Remark that stable configurations are fixed points of the global rule $F$} \cite{FP20}.

Applying these definitions to the particular one-dimensional case of the local rule (2a-FP), relatively to which $\vartheta=2$, a cell $x\in\IZ$ of the configuration $c\in\IN^\IZ$ is \emph{stable} if $c(x)<2$, i.e., if $c(x)\in\parg{0,1}$ is Boolean, and \emph{unstable} if $c(x)\ge 2$. Therefore, in this case a configuration $c$ is \emph{stable} if $\oppA x\in\IZ$, $c(x)\in\parg{0,1}$, and is \emph{unstable} if $\oppE x_0\in\IZ$, $c(x_0)\ge 2$. So, according to Lemma \ref{lm:FP-eq}, a stable configuration is an \emph{equilibrium} state of the induced dynamics.
This means  that if starting from the initial configuration $c_0\in\IN^\IZ$ the dynamical evolution generated by the global rule $F:\IN^\IZ\to\IN^\IZ$ reaches at the time instant $t_0$ an equilibrium configuration $c_{eq}\in\IN^\IZ$, with $F(c_{eq})=c_{eq}$, formally one has the finite sequence of transitions
$c_0\xrightarrow{F} c_1\xrightarrow{F} c_2\xrightarrow{F} \ldots\xrightarrow{F} c_{t_0}=c_{eq}$, where
$F^{t_0}(c)=c_{eq}$, then in the successive dynamical evolution it is $\oppA t\ge {t_0}$, $F^t(c_{eq})=c_{eq}$.
%
\subsection{The particular case of one-dimensional configurations concentrated in the origin}
In this subsection we centered our attention on the particular case of configurations \emph{concentrated} in the origin of the one dimensional lattice of cells $\IZ$, that is of configurations $c:\IZ\to\IN$ such that $c(0)=k$ for a given integer $k\in\IN$ and $c(x)=0$ for any $x\neq 0$.
\begin{proposition}\label{pr:FPequil}
Starting from an initial configuration of the kind $(\bar{0},|k,\bar{0})$, with $k\in\IN$, the unique equilibrium configuration reached after a finite number of time steps of the FP parallel dynamics has one of the two forms:
\begin{enumerate}
\item[(Eq1)]
If $k$ is odd ($=2h+1$) then the final equilibrium configuration is of the symmetric form $(\bar{0},1,\ldots,1,|1,1,\ldots,1,\bar{0})$ centered in the cell $0\in\IZ$ and consisting of a number $2h+1$ of single granules for cell in a \virg{continuous} sequence.
\item[(Eq2)]
If $k$ is even ($=2h$) then the final equilibrium configuration is of the symmetric form $(\bar{0},1,\ldots,1,|0,1,\ldots,1,\bar{0})$  centered in the cell $0\in\IZ$ and consisting of a sequence of $h$ cells with a single granule, followed by a cell whit zero granules in its turn followed by a sequence of $h$ cells with a single granule.
\end{enumerate}
\end{proposition}
\begin{example}\label{ex:FP-020}
The parallel (FP) dynamics of initial state $(\overline{0},|2,\overline{0})$ reaches the equilibrium configuration in a single time step:
\begin{align*}
{}&{} && \qquad c_t
\\
t&=0 &&(\bar{0},0,0,|2,0,0,\bar{0})
\\
t&=1 &&(\bar{0},0,1,|0,1,0,\bar{0})
\end{align*}
\end{example}
\begin{example}\label{ex:FP-030}
Another very simple parallel (FP) dynamics is the one of initial state $(\overline{0},|3,\overline{0})$ whose equilibrium configuration is always reached in a single time step:
\begin{align*}
{}&{} && \qquad c_t
\\
t&=0 &&(\bar{0},0,0,|3,0,0,\bar{0})
\\
t&=1 &&(\bar{0},0,1,|1,1,0,\bar{0})
\end{align*}
This result as consequence of the following triplet transitions $0,0,3\xrightarrow{(SFP2)}*,1,*$, and $0,3,0\xrightarrow{(SFP5)}*,1,*$, and $3,0,0\xrightarrow{(SFP3)}*,1,*$.
\end{example}

We shall discuss now another interesting example of one-dimensional parallel dynamical evolution generated by the local rule (2a-FP), whose initial configuration at time $t=0$ is $c_0=(\overline{0},|6,\overline{0})$ and in agrement with the just stated general result reaches after 8 iterations the equilibrium configuration $c_8=(\bar{0},1,1,1|0,1,1,1,\bar{0})$.
\begin{example}\label{ex:FP-060}
Let us take in examination the configuration $c_0=(\bar{0}|6,\bar{0})$ in the state space $\IN^\IZ$ and let us calculate the parallel dynamics generated by the local rule (2a-FP) starting from it as initial state.
\begin{align*}
{}&{} && \qquad\qquad c_t
\\
t&=0 &&(\bar{0},0,0,0|6,0,0,0,\bar{0})
\\
t&=1 &&(\bar{0},0,0,1|4,1,0,0,\bar{0})
\\
t&=2 &&(\bar{0},0,0,2|2,2,0,0,\bar{0})
\\
t&=3 &&(\bar{0},0,1,1|2,1,1,0,\bar{0})
\\
t&=4 &&(\bar{0},0,1,2|0,2,1,0,\bar{0})
\\
t&=5 &&(\bar{0},0,2,0|2,0,2,0,\bar{0})
\\
t&=6 &&(\bar{0},1,0,2|0,2,0,1,\bar{0})
\\
t&=7 &&(\bar{0},1,1,0|2,0,1,1,\bar{0})
\\
t&=8 &&(\bar{0},1,1,1|0,1,1,1,\bar{0})
\end{align*}
Each transition $t\to t+1$ is obtained by the application of some of the previously discussed 8 cases, where in any of these transitions it is involved the sub-triplet transition $0,0,0\xrightarrow{(SFP1)} *,0,*$. Let us discuss some (not all) of these transitions.
\\
-- $t=0\to t=1$ transition. The first transition $(\bar{0},0,0,0|6,0,0,0,\bar{0})\to (\bar{0},0,0,1|4,1,0,0,\bar{0})$ is the result of the following transitions on sub-triplets: $0,0,6\xrightarrow{(SFP2)} *,1,*$, and $0,6,0\xrightarrow{(SFP5)} *,4,* $, and
$6,0,0\xrightarrow{(SFP3)} *,1,*$.
\\
-- $t=1\to t=2$ transition. Analogously the second transition $(\bar{0},0,0,1|4,1,0,0,\bar{0})\to (\bar{0},0,0,2|2,2,0,0,\bar{0})$ is obtained by the sub-triplets transitions: $0,0,1\xrightarrow{(SFP1)} *,0,*$, and $0,1,4\xrightarrow{(SFP2)} *,2,*$, and
$1,4,1\xrightarrow{(SFP5} *,2,*$, and $4,1,0\xrightarrow{(SFP3)} *,2,*$, and $1,0,0\xrightarrow{(SFP1)} *,0,*$.
And so on for all the other transitions.

The configuration $c_8=(\bar{0},1,1,1|0,1,1,1,\bar{0})$ is of equilibrium, as expected from the general theory since any of its sub-triplet trivially produces the local transition $c(x-1),c(x),c(x+1) \to *,c(x),*$, as consequence of the fact that it is always involved the case (SFP1).

Let us stress that this global parallel dynamics generated by the local rule (2a-FP) starting from the initial configuration $c=(\bar{0}|6,\bar{0})$ cannot be confused with the global dynamics generated by the local rule of height difference (2-a) of section \ref{sec:da-c-a-h}, formally analogous to the (2a-FP), since in this last case of height differences the initial configuration is  $h=(\bar{0},-6|6,\bar{0})$.
\end{example}

From these examples we can induce the following general result, which in any case can be proved.
\begin{proposition}
Let us consider the symmetric configuration $c_0=(\overline{0},|k,\overline{0})$, centered in the origin $x=0$ of the one-dimensional lattice $\IZ$, then the dynamical evolution generated by the parallel application of the FP local rule (2a-FP), $\oppA t\in\IN$, $c_t=F^t(c_0)\in\IN^\IZ$, consists of configurations which are always symmetric and centered in the origin, i.e.,
$$
\oppA t\in\IN,\;\;\oppA x\in\IZ,\quad c_t(-x) =c_t(x).
$$
\end{proposition}

Let us now analyse some of the previous parallel transitions from the point of view of the possible \emph{sequential} application of local rules of the following three different  types:

\begin{enumerate}[(S{I}P1)]
\item
Vertical local rule from left to right, typical of sandpiles,
\\
(VR)$_d$ If $c(x)-c(x+1)\ge 2$, then $c'(x)=c(x)-1$ and $c'(x+1) = c(x+1)+1$.
\\
Vertical local rule from right to left, dual of the previous and typical of symmetric sandpiles of \cite{FMP07},
\\
(VR)$_s$ If $c(x)-c(x-1)\ge 2$, then $c'(x)=c(x)-1$ and $c'(x-1) = c(x-1)+1$.
-- Icepile local horizontal rule, of granules flow from left to right,
\\
(HR)$_d$ If $c(x)=c(x+1)+1$, then $c'(x)=c(x)-1$, and $c'(x+1)=c(x+1)+1$, i.e., under this condition we have the transition $c(x),c(x)-1\to c(x)-1,c(x)$.
\begin{figure}[h!]
\begin{center}
   \includegraphics[width=7cm]{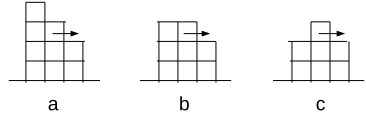}
\end{center}
\vspace{-1cm}
      \caption{The three cases of horizontal rule HR$_d$ with the involvement of the possible state of the cell $x-1$.}
      \label{fig:HR-d}
\end{figure}

In the following we will frequently have to do with the condition $c(x) = 1$ and $c (x + 1) = 0$, or with the transition $1,0\to 0,1$. If now under this condition we consider the possible Boolean state $c (x-1)\in\parg{0,1}$ of the cell at the $x-1$ site we have the following two possible transitions $1,1,0\to 1,0,1$ and $0,1,0\to 0,0,1$.
But even if these are in principle possible transitions we will conventionally consider them as forbidden for reasons of comparison of the digraph of the sequential updating procedure with the parallel dynamics.
\\ \\
\item Icepile local horizontal rule, of granules flow from right to left,
\\
(HR)$_s$ If $c(x)=c(x-1)+1$, then $c'(x)=c(x)-1$ and $c'(x-1)=c(x-1)+1$, corresponding to the transition $c(x)-1,c(x)\to c(x),c(x)-1$.

Analogously to the previous case (HR)$_d$, it will be of great interest the Boolean situation $c(x)=1$ and $c(x-1)=0$, corresponding to the transition $0,1\to 1,0$, with the involvement of the Boolean state of the cell at site $x$ and the two possible transitions $0,1,1\to 1,0,1$ and $0,1,0\to 1,0,0$. Also in this case we conventionally assume them as forbidden.
\end{enumerate}
Summarizing, in the sequel we adopt the following
\begin{description}
\item[Convention (HR)]
In the description of the FP sequential dynamics the two transitions $0,1,0\to 0,0,1$ and $0,1,0\to 1,0,0$ are forbidden in order to compare it with the corresponding parallel dynamics.
\end{description}
%
\begin{enumerate}
\item[(SIP3)]
Bottom-up jump of a granule from left to right
of one height,
\\
(BT)$_d$ If $c(x)\ge 2$ and $c(x)=c(x+1)$, then $c'(x)=c(x)-1$ and $c'(x+1)=c(x+1) +1$.
\\ \\
Bottom-up jump of a granule from right to left
of one height,
\\
(BT)$_s$ If $c(x)\ge 2$ and $c(x)=c(x-1)$, then $c'(x)=c(x)-1$ and $c'(x-1)=c(x-1)+1$.

\begin{figure}[h!]
\begin{center}
   \includegraphics[width=7cm]{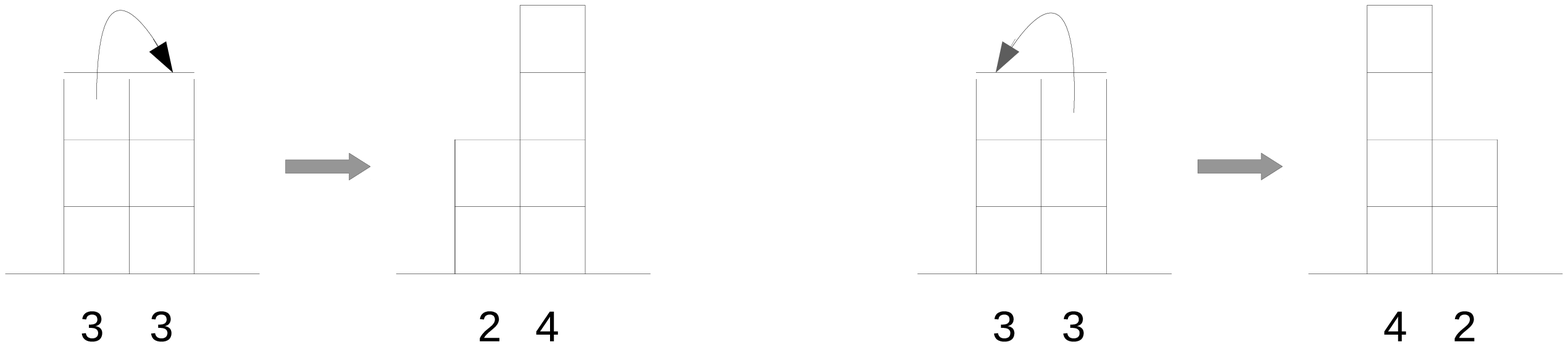}
\end{center}
\vspace{-1cm}
      \caption{Figure at left describes one granule bottom-up jump from left to right, whereas the figure at right describes always a granule bottom-up jump but from right to left.}
      \label{fig:SIP3}
\end{figure}
\end{enumerate}
\begin{description}
\item[Convention (SIP3)]
Since we will consider the sequential update of the pure ice pile model centered in the points (SIP1) and (SIP2) as a theoretical priority, we make the further convention of not using the two updates of the bottom-up jumps (SIP3) until these 
can be considered inessential to reproduce the parallel dynamics.
\\
In other words, we will not use them until we can work without them.
\end{description}
\begin{example}
Let us consider the simplest non-equilibrium configuration $(\overline{0}|2,\overline{0})$ of support centered in the origin of the one-dimensional lattice $\IZ$, and let us apply the local rule (2a-FP) in order to obtain the corresponding equilibrium configuration according to the (one time step) parallel transition
$$
(\overline{0},0,|2,0,\overline{0})\xrightarrow{(PT)}(\overline{0},1,|0,1,\overline{0})
$$

Let us now try to verify in this case the FP claim from \cite{FMP07} that the one-dimensional local rule (2a-FP) is the parallel version of a \emph{pure} sandpile model, i.e., that it can be obtained by the sequential application of the sandpile symmetric (i.e., bidirectional) vertical local rules (VR)$_d$ and (VR)$_s$. The corresponding dynamical digraph is depicted below.
$$
\xymatrix{
{}&0,0,|2,0,0\ar[dr]^{(VR)_d}\ar[dl]_{(VR)_s}&{}
\\
0,1,|1,0,0&{}&0,0,|1,1,0
}
$$
We can therefore establish the following
\begin{description}
\item[Conclusion FP1]
Adopting the convention (HR) the two configurations $(\overline{0},1,|1,0,\overline{0})$ and $(\overline{0},0,|1,1,\overline{0})$ are of equilibrium of the pure sequential sandpile model, but none of them coincide with the expected equilibrium configuration of the parallel transition (PT) seen above: $(\overline{0},1,|0,1,\overline{0})$.
\end{description}
$$
\xymatrix{
{}&0,0,|2,0,0\ar[dr]^{(VR)_d}\ar[dl]_{(VR)_s}\ar[d]^{(PT)}&{}
\\
0,1,|1,0,0&0,1|0,1,0&0,0,|1,1,0
}
$$
\end{example}
\begin{example}
In this example we consider the configuration $(\overline{0},0,|3,0,\overline{0})$ as initial state of the FP parallel dynamics induced from the local rule (2a-FP). Trivially, one gets that the equilibrium configuration of this parallel dynamics is reached in one time step according to the following transformation:
$$
(\overline{0},0,|3,0,\overline{0})\xrightarrow{(PT)}(\overline{0},1,|1,1,\overline{0})
$$
In this particular example the just obtained parallel equilibrium configuration
is reached by the corresponding sequential dynamics based on the unique vertical rule of the pure sandpile model, as shown in the following digraph:
$$
\xymatrix{
{}&0,0,|3,0,0\ar[dr]^{(VR)_d}\ar[dl]_{(VR)_s}\ar[dd]^{(PT)}&{}
\\
0,1,|2,0,0\ar[dr]_{(VR)_d}&{}&0,0,|2,1,0\ar[dl]^{(VR)_s}
\\
{}&0,1,|1,1,0&{}
}
$$

Therefore, from this particular example it might appear correct the FP's claim that the model based on the parallel application of the local rule (2) referred to in the Introduction and characterizing their article, or its peculiar one-dimensional version (2a), is about sandpiles.
Indeed, in this example the parallel equilibrium configuration is obtained through appropriate sequential applications of the vertical rules from point (SIP1) only, albeit bilateral, which characterize the sandpile dynamics.

In any case, below we still want to describe the sequential dynamics of the pure model of bilateral icepiles starting from the same initial configuration according to the vertical (SIP1) and horizontal (SIP2) rules introduced above, rather than a bidirectional pure sandpile model centered in the unique vertical (SIP1) rule. The corresponding sequential dynamics is drawn in the following digraph:
$$
\xymatrix{
{}&0,0,|3,0,0\ar[dr]^{(VR)_d}\ar[dl]_{(VR)_s}\ar[dd]^{(PT)}&{}
\\
0,1,|2,0,0\ar[dr]_{(VR)_d}\ar[d]_{(HR)_s}&{}&0,0,|2,1,0\ar[dl]^{(VR)_s}\ar[d]^{(HR)_d}
\\
0,2,|1,0,0\ar[d]_{(VR)_s}&0,1,|1,1,0&0,0,|1,2,0\ar[d]^{(VR)_d}
\\
1,1,|1,0,0&{}&0,0,|1,1,1
}
$$

In this case, and under the (HR) convention, we have three equilibrium configurations, one of the parallel update $(\overline{0},0,1,|1,1,0,\overline{0})$ and other two as results of the sequential update $(\overline{0},1,1,|1,0,0,\overline{0})$ and $(\overline{0},0,0,|1,1,1,\overline{0})$.
\end{example}
\begin{remark}
This particular result of the presence of three equilibrium configurations of the sequential FP dynamics suggest some interesting consideration about their \virg{physical} symmetry.

First of all, for any fixed integer $a\in\IZ$ let us introduce the so-called $a$--\emph{translation} (also, $a$--\emph{left shift})
operator on the configuration space $\IN^\IZ$, denoted as $T_a:\IN^\IZ\to\IN^\IZ$ and defined by the correspondence
$(\ldots,c(-1),|c(0),c(1),\ldots) \xrightarrow{\;T_a\;} (\ldots,c(a-1),|c(a),c(a+1),\ldots)$.

The collection of all such translations, $\mathcal{T}(\IN^\IZ)=\parg{T_a:a\in\IZ}$, has a structure of abelian group with respect to the operation of composition, $T_a\circ T_b =T_b\circ T_a=T_{a+b}$. In particular we have that
the neutral element is the \emph{identical} translation $T_0$ ($\oppA c\in\IN^\IZ$, $T_0(c)=c$) since $T_0\circ T_a=T_a\circ T_0= T_a$, for every $T_a$,
and the inverse of a generic translation $T_a$ is the translation $(T_a)^{-1}=T_{-a}$ since $T_a\circ T_{-a}=T_0=T_{-a}\circ T_a$.

Now on the sate space of all configurations $\IN^\IZ$ the following is an equivalence relation
$$
\text{Let}\; c_1,c_2\in\IN^\IZ,\quad\text{then}\;\; c_1\sim c_2\;\;\text{iff}\;\; \oppE a\in\IZ\;s.t.\; c_1 = T_a c_2.
$$
That is, two configurations are mutually equivalent iff one is obtained from the other by a suitable translation, and the configuration space can be decomposed by the collection of all pairwise disjoint nonempty equivalence classes relatively to translations $[c]_\sim :=\parg{c':c'\sim c}$.
\\
From the physical point of view the abelian group $\mathcal{T}(\IN^\IZ)$ of all translations of the configuration space is a \emph{symmetry} of this space, and so two configurations $c_1\sim c_2$ are \emph{equivalent} by the symmetry of translation.

In particular $(\overline{0},1,0,|1,0,0,\overline{0}) =T_1(\overline{0},0,1,|0,1,0,\overline{0})$ and $(\overline{0},0,0,|1,0,1,\overline{0})=T_{-1}(\overline{0},0,1,|0,1,0,\overline{0})$, and so the three equilibrium configurations of the sequential dynamics are mutually equivalent among them relatively to translations, in other words, they belong to the same translation equivalence class.
\end{remark}
\begin{example}
The further example we will now consider is based on the configuration consisting of four granules centered at the origin, $(\overline{0},0,|4,0,\overline{0})$, considered as the initial state of the following FP dynamics generated by the local rule (2a-FP) reaching the equilibrium configuration $(\overline{0},1,1,|0,1,1,\overline{0})$ after four time steps:
\begin{align*}
{}&{} &  c_t\phantom{400000}
\\
t&=0 &(\bar{0},0,0|4,0,0,\bar{0})
\\
t&=1 &(\bar{0},0,1|2,1,0,\bar{0})
\\
t&=2 &(\bar{0},0,2|0,2,0,\bar{0})
\\
t&=3 &(\bar{0},1,0|2,0,1,\bar{0})
\\
t&=4 &(\bar{0},1,1|0,1,1,\bar{0})
\end{align*}
The digraph of the sequential updating procedure obtained by the use of the pure local rule (2a-FP) characterizing the one-dimensional sandpiles, would seem to be the one drawn below, where we neglect all the sequential transitions that generate orbits that in any case lead to equilibrium configurations different from the required \virg{parallel} one $(\overline{0},1,1,|0,1,1,\overline{0})$.
$$
\xymatrix{
{}&0,0,|4,0,0\ar[dl]_{(VR)_s}\ar[dd]^{(PT)}\ar[dr]^{(VR)_d}&{}
\\
0,1,|3,0,0\ar[dr]_{(VR)_d}\ar@{-->}[d] &{}& 0,0,|3,1,0\ar[dl]^{(VR)_s}\ar@{-->}[d]
\\
0,2,|2,0,0&0,1,|2,1,0\ar[dd]^{(PT)}&0,0,|2,2,0
\\
{}&{}&{}
\\
{}&0,2,|0,2,0\ar[dl]_{(VR)_s}\ar[d]\ar[dr]^{(VR)_d}&{}
\\
1,1,|0,2,0\ar[dr]_{(VR)_d}&1,0|2,0,1\ar[d]&0,2,|0,1,1\ar[dl]^{(VR)_s}
\\
{}&1,1,|0,1,1&{}
}
$$

The only negative point of this digraph lies in the parallel transition $(\overline{0},0,1,|2,1,0,\overline{0})\xrightarrow{(PT)}(\overline{0},0,2,|0,2,0,\overline{0})$ which cannot be justified by the sequential application of the vertical rule (SIP1) since the input configuration $c_1=(\overline{0},0,1,|2,1,0,\overline{0})$ does not present any at least 2 \virg{critical jump} of the height differences (indeed, for any $x\in\IZ$, $c_1(x)-c_1(x+1)\le 1$).

On the contrary, this result can necessarily be obtained by the appropriate application of both horizontal (SIP2) and bottom-up jump (SIP3) local rules, as shown in the partial diagram below which completes the overall dynamics seen above:
$$
\xymatrix{
{}&0,1,|2,1,0\ar[dl]_{(HR)_s}\ar[dd]^{(PT)}\ar[dr]^{(HR)_d}&{}
\\
0,2,|1,1,0\ar[dr]_{(BT)_d}&{}&0,1,|1,2,0\ar[dl]^{(BT)_s}
\\
{}& 0,2,|0,2,0&{}
}
$$
\end{example}
\begin{description}
\item[Conclusion FP2]
The present example shows that the expected equilibrium configuration of the parallel dynamics generated by the one-dimensional local rule (2a-FP) of the Formenti-Perrot (FP) model is obtained not only through the sequential use of only the vertical rule (SIP1) (this rule is not sufficient to obtain the expected goal) but by the necessary intervention of the horizontal rule (SIP2) plus the relevant use of the bottom-up jump rule (SIP3). And this is what we referred to in the Introduction as the \emph{spurious} icepile model.
\\
This means that the FP claim in \cite{FP20} that theirs is a model of a sandpiles dynamic is not correct since in order to obtain the expected equilibrium parallel result, at least in this simple initial configuration of the total number of $N = 4$ granules centered in the origin, all the rules of a \virg{spurious} icepile model \emph{must} necessarily be sequentially applied.
\\
Therefore the title of their article, and the whole section 2.1, which both explicitly refer to sandpiles is incorrect because at least they should refer to \emph{spurious icepile} model, as this counterexample shows.
\end{description}
\begin{example}\label{ex:FP-060-seq}
Let us analyse some transitions of the parallel dynamics generated by the one-dimensional local rule (2a-FP) of the Formenti--Perrot model discussed in example \ref{ex:FP-060}, starting from the initial state $(\bar{0}|6,\bar{0})$, as the results of the sequential application of the previously discussed three types of local rules.
\\
- The parallel transition (PT) from $t=0$ to $t=1$ can be decomposed by two sequential transitions, each consisting of two steps. Firs of all, let us draw the corresponding two steps conventional sequential digraph
$$
\xymatrix{
{}&0,0|6,0,0\ar[dl]_{(VR)_s}\ar[dr]^{(VR_d)}\ar[dd]^{(PT)}&{}
\\
0,1|5,0,0\ar[dr]^{(VR_d)}\ar[d]_{(VR)_s} &{}&0,0|5,1,0\ar[dl]_{(VR)_s}\ar[d]^{(VR)_d}
\\
0,2|4,0,0&0,1|4,1,0&0,0|4,2,0
}
$$
Then, we have the following two paths towards the parallel configuration at time $t=1$:
\begin{align*}
&(\bar{0},0,|6,0,\bar{0})\xrightarrow{\text{(VR)$_s$}} (\bar{0},1,|5,0,\bar{0})\xrightarrow{\text{(VR)$_d$}} (\bar{0},1,|4,1,\bar{0})
\\
&(\bar{0},0,|6,0,\bar{0})\xrightarrow{\text{(VR)$_d$}} (\bar{0},0,|5,1,\bar{0})\xrightarrow{\text{(VR)$_s$}} (\bar{0},1,|4,1,\bar{0})
\end{align*}
- Neglecting the \virg{secondary} sequential transitions (dashed arrows) with respect to obtaining the parallel ones, below we draw the transitions from $t=1$ to $t=2$, and from the latter to $t=3$, all involving the sequential vertical transitions (VR).
$$
\xymatrix{
{}&0,1|4,1,0\ar[dl]_{(VR)_s}\ar[dr]^{(VR_d)}\ar[dd]^{(PT)}&{}
\\
0,2|3,1,0\ar[dr]^{(VR)_d}\ar@{-->}[d] &{}&0,1|3,2,0\ar[dl]_{(VR)_s}\ar@{-->}[d]
\\
{}&0,2|2,2,0\ar[dr]_{(VR)_d}\ar[dl]^{(VR)_s}\ar[dd]^{(PT)}&{}
\\
1,1,|2,2,0\ar[dr]_{(VR)_d}\ar@{-->}[d]&{}&0,2,|2,1,1\ar[dl]^{(VR)_s}\ar@{-->}[d]
\\
{}&1,1,|2,1,1&{}
}
$$
- More interesting is the drawn below parallel transition (PT) from time $t=3$ to time $t=4$, starting from the configuration at time $t=3$ as initial state, whose \virg{essential} sequential digraph (i.e., neglecting the \virg{secondary} transitions) necessarily involves, besides the (HR) horizontal transitions, also the (BT) transitions of bottom-to-top jumps of a granule.
$$
\xymatrix{
{}&1,1,|2,1,1\ar[dl]_{(HR)_s}\ar[dr]^{(HR)_d}\ar[dd]_{(PT)}&{}
\\
1,2,|1,1,1\ar[dr]_{(BT)_d}&{}&1,1,|1,2,1\ar[dl]^{(BT)_s}
\\
{}&1,2|0,2,1&{}
}
$$
The state at time $t=4$ of the parallel transition is reached by the following two sequential paths in which a relevant role is played by the two jumps from bottom-to-top (BT)$_d$ and (BT)$_s$.
\begin{align*}
&(\bar{0},1,1,|2,1,1,\bar{0})\xrightarrow{(HR)_s} (\bar{0},1,2,|1,1,1,\bar{0})\xrightarrow{(BT)_d} (\bar{0},1,2,|0,2,1,\bar{0})
\\
&(\bar{0},1,1,|2,1,1,\bar{0})\xrightarrow{(HR)_d} (\bar{0},1,1,|1,2,1,\bar{0})\xrightarrow{(BT)_s} (\bar{0},1,2,|0,2,1,\bar{0})
\end{align*}

Note that from this state onwards, all the essential sequential transitions that justify the parallel transitions of the chain $1,2,|0,2,1\xrightarrow{(PT)}\; 2,0,|2,0,2\xrightarrow{(PT)}\;1,0,2|0,2,0,1\xrightarrow{(PT)}\;1,1,0,|2,0,1,1$ necessarily involve the transitions bottom-top (BT) in an unavoidable point.
\end{example}
\section{Conclusions about the now discussed one-dimensional FP model, open questions and further developments}
As first conclusions we can summarize the main results obtained by the one-dimensional discussion about the FP model widely treated in section \ref{sc:FP-inter} in the following points.
\begin{enumerate}[(Co1)]
\item
The parallel application of the local rule (2a-FP) is not able to produce the \emph{canonical} sandpile dynamics generated by the local rule (VR) (or its equivalent version (VR-a)).
\item
On the contrary, the parallel dynamics generated by (2a-FP) is precisely the one of a \emph{spurious symmetrical icepile} as suitable sequential application of the following three rules:
\begin{enumerate}[(S{I}P1)]
\item
Vertical rules both from left to right (VR)$_d$, than its dual from right to left (VR)$_s$, typical of the symmetric sandpiles of \cite{FMP07}.
\item
Icepile horizontal rules, of a single cell flowing both from left to right (HR)$_d$, than from right to left (HR)$_s$,  in presence of horizontal plateaus.
\item
Jump of a granule from the bottom to the top of a single height, both from left to right (BT)$_d$ than from right to left (BT)$_s$.
\end{enumerate}
\end{enumerate}
\begin{description}
\item[First important conclusion]
In our opinion it turns out to be quite improper to entitle the Formenti-Perrot paper \cite{FP20} with the explicit reference to \virg{sandpiles} on a lattice when really it is modelled the situation of \emph{symmetrical icepiles} with the furthermore involvement of a \emph{spurious} law consisting in unusual (anti-gravitational) jumps of granules towards the top, with the certainty that this approach will never be  able to simulate the parallel version of classical one-dimensional sandpiles governed by the \emph{unique} standard vertical rule (VR), or its equivalent formulation (1a).
\\
This means that in treating this argument one must take in in consideration the terminology of \virg{ice granules} instead of the one of \virg{sand granules}.
\item[Second important conclusion]
The main focus of the paper consists in a comparison of the one-dimensional FP spurious symmetric icepile model, whose sequential version is based on the above three \virg{local rules} (SI1)--(SIP3), with the standard GK sandpile model, which is not symmetric.
\\
This comparison is not at all correct since GK is not symmetric, contrary to the FP model. In order to have a right comparison it is necessary to investigate, and this will be done in some forthcoming papers actually in a draft form, the following symmetric model:
\end{description}
\begin{enumerate}[(SM1)]
\item
\emph{Symmetric sandpile model}. A symmetric model of sandpiles (SSPM) as symmetric version of sandpile model (SPM) is introduced and discussed in \cite{FMP07}, and at the best of our knowledge, it is the unique contribution to this argument one can found in literature. Quoting from \cite{FMP07}: \virg{The new model follows the rules of SPM but it applies them in both directions}. Moreover, the dynamics is the one generated by the sequential updating of the sites, in which \virg{only one grain is allowed to move per time step [...] according to the following guidelines: (i) a grain can move either to the left or to the right, if the [height] difference is more than 2; (ii) when a grain can move only in one direction, it follows the SPM rule (right) or it symmetric (left). [...] The model is intrinsically sequential: only one grain moves at each time step}.

In \cite{CM21} we introduce the \emph{global transition function} as parallel application to any cell of the one dimensional lattice $\IZ$ of the following \emph{symmetric local rule}.
\\
$\oppA c\in\IN^\IZ,\;\oppA x\in\IZ$,
\begin{align*}
c'(x) = c(x)
&+\mathrm{H}\big(c(x)-c(x+1)\big)\Big(\mathrm{H}\big(c(x-1)-c(x)-2 \big) - \mathrm{H}\big( c(x)-c(x+1)-2\big)\Big)
  \\
&+\mathrm{H}\big(c(x)-c(x-1)\big)\Big(-\mathrm{H}\big(c(x)-c(x-1)-2 \big) + \mathrm{H}\big(c(x+1)-c(x)-2 \big)\Big)
\end{align*}

This local rule is \emph{symmetric} in the sense that it formalizes the simultaneous action of the standard GK local rule  in which \virg{a grain of sand tumbles from site $x$ to site $x+1$ [i.e., direction from left to right] if the height difference $c(x)-c(x+1)$ is at least 2} \cite{GK93}
$$
\oppA c\in\IN^\IZ,\;\oppA x\in\IZ,\quad c'(x)= c(x)+\mathrm{H}\Big(c(x-1)-c(x)-2\Big) -\mathrm{H} \Big(c(x)-c(x+1)-2\Big)
$$
and the dual right-to-left local rule in which a grain of sand tumbles from site $x$ to site $x-1$ [i.e., direction from right to left] if the height difference $c(x-1)-c(x)$ is at least 2
$$
\oppA c\in\IN^\IZ,\;\oppA x\in\IZ,\quad c'(x)= c(x)-\mathrm{H}\Big(c(x)-c(x-1)-2\Big) +\mathrm{H} \Big(c(x+1)-c(x)-2\Big)
$$
\end{enumerate}